\documentclass[11pt]{article}
\usepackage[english]{babel}
\usepackage{amsmath,amssymb}
\newcommand{\heffm}{\bar{H}_{\mathrm{eff}}^{\mbox{\tiny $(\!-\!)$}}}
\newcommand{\heffp}{\bar{H}_{\mathrm{eff}}^{\mbox{\tiny $(\!+\!)$}}}
\newcommand{\heffo}{\bar{H}_{\mathrm{eff}}^{\mbox{\tiny $(0)$}}}
\newcommand{\updown}{{\mbox{\tiny $\updownarrow$}}}
\newcommand{\sapprox}{{\mbox{\tiny $\sim\hspace{-1.7mm}\updownarrow$}}}
\newcommand{\gn}{g(\hat{n})}
\newcommand{\hl}{H(\lambda)}
\newcommand{\hlt}{H(\lambda; t)}
\newcommand{\ho}{H_0}
\newcommand{\hi}{H_{\diamond}}
\newcommand{\uo}{U_0}
\newcommand{\uot}{U_0(t)}
\newcommand{\hm}{H_m}
\newcommand{\hn}{H_n}
\newcommand{\hil}{\hi (\lambda)}
\newcommand{\hilt}{\hi (\lambda ;t)}
\newcommand{\lt}{(\lambda ;t)}
\newcommand{\rotev}{T}
\newcommand{\zl}{Z(\lambda)}
\newcommand{\zlt}{Z(\lambda ;t)}
\newcommand{\lto}{(\lambda;t,t_0)}
\newcommand{\clt}{C(\lambda ;t)}
\newcommand{\cl}{C(\lambda)}
\newcommand{\ttt}{\mathfrak{t}}
\newcommand{\cltint}{\!\int_0^t\! C(\lambda ;\ttt)\, \de\ttt}
\newcommand{\rotham}{{\tilde{H}}}
\newcommand{\rothamlt}{\tilde{H}(\lambda ;t)}
\newcommand{\spa}{\!\!\!}
\newcommand{\de}{\mathrm{d}}
\newcommand{\der}{\frac{\mathrm{d}\ }{\mathrm{d}t}}
\newcommand{\Ad}{\mathrm{Ad}}
\newcommand{\ad}{\mathrm{ad}}
\newcommand{\X}{\mathfrak{X}}
\newcommand{\zz}{\mathcal{Z}}

\newcommand{\rrc}{\lfloor C\rfloor}

\newcommand{\cc}{\mathfrak{C}}

\newcommand{\hb}{\overline{H}(\lambda)}
\newcommand{\hG}{\breve{\mathsf{G}}}

\newcommand{\tc}{{^{t}\!C}}
\newcommand{\tauc}{{^{\tau}\!C}}

\newcommand{\enne}{\mathsf{N}}
\newcommand{\enneind}{_{\mbox{\tiny $[\mathsf{N}]$}}}
\newcommand{\infc}{{^{\infty}\hspace{-0.4mm}C}}
\newcommand{\infz}{{^{\infty}\!Z}}
\newcommand{\minc}{{^{\bowtie}\hspace{-0.2mm}C}}
\newcommand{\minz}{{^{\bowtie}\!Z}}
\newcommand{\extra}[1]{\left\vert\rangle#1\langle\right\vert_{\ho}}
\newcommand{\infra}[1]{\left\langle\vert#1\vert\right\rangle_{\!\ho}}

\newcommand{\Extra}[1]{\left[\vert\rangle#1\langle\vert\right]_{\ho}}
\newcommand{\GG}{\mathcal{G}}
\newcommand{\GGG}{\mathsf{G}}
\newcommand{\RR}{\mathcal{R}}
\newcommand{\RRR}{\mathsf{R}}
\newcommand{\zzzl}{\mathfrak{Z}_{\mbox{\tiny $\mathsf{L}$}}}
\newcommand{\zzzr}{\mathfrak{Z}_{\mbox{\tiny $\mathsf{R}$}}}
\newcommand{\period}{\mathtt{T}}

\newcommand{\prim}{{\smallint\hspace{-1.2mm} F}}

\newcommand{\mean}[1]{{\{\hspace{-1.3mm}\{\hspace{0.4mm}#1\hspace{0.5mm}\}\hspace{-1.3mm}\}}}
\newcommand{\mmean}[1]{{\big\{\hspace{-1.6mm}\big\{\hspace{0.5mm}#1\hspace{0.5mm}\big\}\hspace{-1.6mm}\big\}}}
\newcommand{\primitive}{{\smallint\hspace{-2.6mm}\mbox{--}\hspace{-0.2mm}}}
\newcommand{\app}{\stackrel{\ \lambda^\enne}{\approx}}
\newcommand{\appr}{\,\stackrel{\lambda}{\approx}\,}
\newcommand{\af}{{a_{f}^{\phantom{\dagger}}}}
\newcommand{\afd}{{a_{f}^{\dagger}}}
\newcommand{\eif}{e^{i\phi}}
\newcommand{\emif}{e^{-i\phi}}
\newcommand{\fun}{\Phi}
\newcommand{\avxp}{\mathrm{avxp}}
\newcommand{\idc}{\mathrm{Id}_{\mbox{\tiny $\mathbb{C}^2$}}}
\newcommand{\idf}{\mathrm{Id}_{\mbox{\tiny $\mathcal{H}_{\mathrm{F}}$}}}

\newcommand{\effe}{\mathsf{f}}
\newcommand{\TC}{T_C}
\newcommand{\TZ}{T_Z}
\newcommand{\eff}{\tilde{E}}
\newcommand{\wuuno}{V_{\!\mbox{\tiny $(1)$}}}
\newcommand{\wudue}{V_{\!\mbox{\tiny $(2)$}}}

\title{A new perturbative expansion of the time evolution operator
associated with a quantum system
\vspace{0.5cm} }

\author{
{\Large P. Aniello\,}\footnote{\; Paolo.Aniello@na.infn.it} \vspace{0.3cm}\\
{\small Istituto Nazionale di Fisica Nucleare, Sezione di
Napoli,}\\{\small and}\\
{\small Dipartimento di Scienze Fisiche dell'Universit\`{a} di
Napoli \lq\lq Federico II\rq\rq,}
\\
{\small Complesso Universitario di Monte S. Angelo, Via Cintia -
80126 Napoli, Italy} \vspace{0.4cm}}

\begin{document}

\maketitle
\begin{abstract} \noindent
{\sf A novel expansion of the evolution operator associated with
a  -- in general, time-dependent -- perturbed quantum Hamiltonian is presented.
It is shown that it has a wide range of possible realizations
that can be fitted according to computational convenience or to satisfy
specific requirements. As a remarkable example, the quantum Hamiltonian
describing a laser-driven trapped ion is studied in detail.}
\end{abstract}



\section{Introduction}

The explicit determination of the evolution operator associated
with a quantum system --- namely, the determination of its
explicit action on the state vectors, or, equivalently, on a given
orthonormal basis --- is, in general, a `touchy business'. If the
Hamiltonian of the system does not depend on time and has the form
of the sum of a solvable unperturbed Hamiltonian plus an analytic
perturbation, one can use the tools of standard perturbation
theory~{\cite{Kato-pert} for linear operators, based on the
expansion of the resolvent, in order to get approximate
expressions of the evolution operator. Alternatively, one can
apply a suitable operator perturbative approach~\cite{Aniello1,
Aniello2, AnielloH, Militello, Aniello3} in order to obtain a very
convenient perturbative expansion of the evolution operator whose
truncations have the remarkable property of
forming one-parameter groups of unitary operators. The
main goal of this paper is to extend, in a natural way, such a
perturbative approach to the general case where the quantum
Hamiltonian may depend on time.

We stress  that if a quantum Hamiltonian is time-dependent
--- i.e.\ it describes a non-isolated quantum system --- the
task of determining the associated time evolution operator
is a tough problem since, whenever the values
of the Hamiltonian at different times do not commute, the
evolution operator does not admit a simple formal expression.\\ In
two fundamental papers~\cite{Dyson}, Dyson developed an expansion
of the evolution operator that has been adopted extensively in any
field of physics. Dyson expansion has a transparent physical
interpretation in terms of time ordered elementary processes which
makes its application particularly appealing, especially in
quantum field theory. On the other hand, for many applications,
Dyson expansion has severe drawbacks, as a low convergence rate
and the lack of unitarity of its truncations~\cite{Pechukas}.\\
Later, Magnus~\cite{Magnus} introduced an expansion of the
evolution operator such that each of its truncations retains the
property of being unitary. Magnus expansion has been
`rediscovered' and re-elaborated several times, and applied
successfully to several problems (see~\cite{Moan} and references
therein). It is written in the form of the exponential of the
expansion of a suitable time-dependent anti-hermitian operator
which can be deduced, order by order, from the Hamiltonian of the
system. Now, precisely for this reason --- just like for the
evolution operator generated by a time-independent Hamiltonian ---
the problem of {\it computing} explicitly the action of (any
truncation of) the Magnus expansion on the state vectors is
non-trivial. Truncating the power expansion of the exponential
would lead to non-unitary expressions, thus to the loss of the
most important feature of Magnus expansion. Besides, in the
important case of a time-independent Hamiltonian, there is no link
between Rayleigh-Schr\"odinger-Kato perturbation theory for linear
operators and Magnus expansion, a clue suggesting that the power
of the perturbative approach is not `fully exploited' by this
expansion.\\ Then, the issue of finding a generalization of Magnus
expansion retaining the property of having unitary truncations,
but allowing more convenient solutions
--- or solutions having special properties --- arises in a natural
way. In the present paper, we have tried to achieve this result.
We introduce a perturbative decomposition of the evolution
operator that generalizes Magnus expansion, and opens the
possibility of obtaining computational advantages and of
satisfying specific requirements for the perturbative solutions by
suitably fixing, at each perturbative order, certain arbitrary
operators or operator-valued functions. It is important to remark
that, in the case where the Hamiltonian does not depend on time,
our decomposition, differently from Magnus expansion, assumes a
special meaning and has a precise link with
standard perturbation theory for linear operators.\\
The idea of generalizing Magnus expansion is not
completely  new. It appears in a paper by Casas~{\it et
al.}~\cite{Casas} in which the authors introduce the {\it
Floquet-Magnus expansion} for the evolution operator associated
with a (interaction picture) Hamiltonian depending periodically on
time. However, it turns out that our approach generalizes the one
proposed by Casas~{\it et al.}\ even in the case where the
interaction picture Hamiltonian is periodic on time.

We have made the choice of skipping, as far as possible,
mathematical complications. Our choice is motivated by various
reasons. First, we believe that heuristic investigation should
always precede rigorous re-elaboration. Once that it is clear what
the basic `rules of the game' are, one can adopt the most
appropriate mathematical tools. Moreover, we avoid the risk of
hiding in a thick cloud of technicalities the main ideas and of
discouraging those who may want to {\it apply} our method for
solving problems. It should be also observed that a recent trend
in quantum mechanics is to focus on systems which can be described
by effective Hamiltonians in finite-dimensional Hilbert spaces
(consider the huge research area related to quantum computation
and quantum information theory; see ref.~\cite{Galindo} and the
rich bibliography therein). The study of these systems is not
affected by the technicalities associated with the
infinite-dimensional spaces but retains all the most intriguing
features of quantum physics. \\
The structure of the paper is the following. We begin introducing,
in sect.~{\ref{ion}}, before going through the details of a our
perturbative approach, a significant example --- the quantum
Hamiltonian describing a laser-driven trapped ion --- example that
will play a double role: a starting motivation for our analysis
and a concrete playing field for practising our theoretical results. In
sect.~{\ref{basic}} we establish the general form of the
perturbative decomposition of the evolution operator that we
propose. In the subsequent two sections, we pursue the task of
finding a recursive procedure that allows to compute, order by
order, the various terms of our perturbative expansion. Precisely: in
sect.~{\ref{independent}} we study the case where the Hamiltonian
does not depend on time and investigate the link with standard
perturbation theory for linear operators; in
sect.~{\ref{generalcase}} we consider the general case of a
time-dependent Hamiltonian. Next, sect.~{\ref{applications}} is
devoted to study in detail the significant example introduced in sect.~{\ref{ion}},
in such a way to illustrate the main
features of our perturbative expansion of the evolution operator.
Finally, in sect.~{\ref{conclusions}}, conclusions are drawn, with
a quick glance to further applications.


\section{A remarkable example: the ion trap Hamiltonian}
\label{ion}

Before introducing our perturbative approach, we think that is worth considering
first a remarkable example: the Hamiltonian describing a one-dimensional
laser-driven ion trap. The study of this Hamiltonian, which will be
performed systematically in sect.~{\ref{applications}}, allows to illustrate
in a simple way all the main points of the theory developed in the subsequent sections.

A two-level ion of mass $\mu$ in a potential trap, with strong
confinement along the $y$ and $z$ axes, and weak harmonic binding
of frequency $\nu$ along the $x$-axis (the `trap axis'), can be
described --- neglecting the motion of the ions transverse to the
trap axis --- by a Hamiltonian of the following type ($\hbar=1$):
\[
H_0=\nu \, \hat{n} +\frac{1}{2}\,\epsilon\,\sigma_z\, ,
\]
where $\hat{n}=a^\dagger a$ is the `number operator' --- with $a$ denoting
the vibrational annihilation operator
\[
a=\left(\frac{\mu\,\nu}{2}\right)^{\frac{1}{2}}
\left(\hat{x}+\frac{i}{\mu\,\nu}\,\hat{p}_x\right)
\]
---
and $\sigma_z$ the effective spin operator associated with the
internal degrees of freedom of the ion. Let us suppose now that
the ion is addressed by a laser beam of frequency $\alpha$ in a
`traveling wave configuration'. Then, the Hamiltonian describing
the physical system becomes (see, for instance, ref.~{\cite{Cirac}}):
\begin{equation} \label{initialham}
H_0 + H_\updown(t)\, ,
\end{equation}
where the time-dependent interaction term $H_\updown (t)$ is
defined by
\begin{equation} \label{interm}
H_\updown (t) := \Omega_{R}\left( e^{i\alpha t}\,
D(i\eta)^{\dagger}\,\sigma_- + e^{-i\alpha t}\, D(i\eta)\,\sigma_+\right),
\end{equation}
with $\Omega_{R}$ denoting the Rabi frequency (which is
proportional to the intensity of the laser field) and with
$\sigma_{\pm}:=\left|\pm\right\rangle\left\langle\mp\right|\,$,\,
$\sigma_+=\sigma_-^\dagger\ $
($\sigma_z\left|\pm\right\rangle=\pm\left|\pm\right\rangle$).
Moreover, we have set:
\begin{equation}
D(i\eta):=\exp \left(i \eta \left( a + a^{\dagger}\right) \right),
\end{equation}
where
\begin{equation}
\eta := \frac{k_L\cos\phi}{\sqrt{2\mu\nu}}
\end{equation}
--- with $\mathbf{k}_L$ the wavevector and $\phi$ the angle between the
$x$-axis and $\mathbf{k_L}$ --- is the so-called `Lamb-Dicke
parameter'. In the case where $\eta\ll 1$ (`Lamb-Dicke regime')
--- a case often occurring in applications\footnote{Notice that in
a trap with a {\it linear geometry}, like the one we are
considering here, one can modify the incidence angle $\phi$ of the
laser beam with respect to the principal axis of the trap in order
to control the Lamb-Dicke parameter $\eta$.}
--- one can keep only those terms in the power expansion
of $D(i\eta)$ which are at most linear in $\eta$:
\begin{equation} \label{appham}
H_\updown(t)\approx \Omega_R \left(e^{i\alpha
t}\,(1-i\eta(a+a^\dagger))\,\sigma_-+\, h.c.\right)=: H_\sapprox
(t)\, .
\end{equation}

Observe that the problem of dealing with the time-dependent
Hamiltonian $H_0 + H_\updown(t)$ can be bypassed by switching
to the interaction picture with reference Hamiltonian
$\frac{1}{2}\,\alpha \,\sigma_{z}$. Indeed, setting
\begin{equation}
R_t^{} := \exp\left(\!-\frac{i}{2}\,\alpha \,\sigma_z\, t \right),
\end{equation}
one obtains the time-independent `rotating frame Hamiltonian'
\begin{eqnarray}
R_{t}^{\dagger}\left(H_0 +
H_\updown(t)-\frac{1}{2}\, \alpha\,\sigma_{z}\right)
R_{t}^{}  \spa
&=& \spa \nu\,\hat{n} +\frac{1}{2}\,\delta\,
\sigma_{z}+\Omega_{R}\left( D(i\eta)^{\dagger}\,\sigma_- +
D(i\eta)\, \sigma_+ \right) \nonumber \\ \mbox{(Lamb-Dicke
regime: $\eta\ll 1$)}\ \ \ \ & \approx & \spa \Omega_R
\left((1-i\eta(a+a^\dagger))\,\sigma_- +\, h.c.\right) ,
\end{eqnarray}
where $\delta:=\epsilon-\alpha$ is the ion-laser detuning. In many
applications, the condition $\Omega_R\ll\nu$ is also verified;
hence, it is natural to introduce the (dimensionless) perturbative
parameter
\begin{equation}
\lambda\equiv \frac{\Omega_R}{\nu}\ll 1\,.
\end{equation}
Thus, the study of the time-dependent
Hamiltonian~{(\ref{initialham})} can ultimately be reduced to the
study of the simpler time-independent Hamiltonian
\begin{equation} \label{simpler}
\bar{H}:=\nu\,\hat{n} +\frac{1}{2}\,\delta\, \sigma_{z}+\lambda\,\nu
\left((1-i\eta(a+a^\dagger))\,\sigma_- +\, h.c.\right) ,
\end{equation}
which has the form of a trivially solvable Hamiltonian\footnote{Namely,
a Hamiltonian such that a complete orthonormal set of eigenvectors is
given simply by the standard basis $\{|n\rangle\otimes|\pm\rangle\}$.}
$\bar{H}_0:= \nu\,\hat{n} +\frac{1}{2}\,\delta\, \sigma_{z}$
plus a `small perturbation'
$\bar{H}_\updown:= \lambda\,\nu
\left((1-i\eta(a+a^\dagger))\,\sigma_- +\, h.c.\right)$  .

The exact eigenvalues and eigenprojectors of the Hamiltonian $\bar{H}$,
despite its formal simplicity, are not known. Hence,
at this point, it would seem reasonable to adopt a (time-independent)
perturbative approach for studying the Hamiltonian $\bar{H}$.
However, this is not what it is usually done in the literature.
In fact, with the aim of providing an approximate expression of the
evolution operator associated with the Hamiltonian
$\bar{H}$, the {\it rotating wave approximation} (RWA)
--- see, for instance, ref.~{\cite{Schleich}} --- is
usually applied. This approximation amounts to passing to
a further interaction picture with reference
Hamiltonian $\bar{H}_0$, so obtaining the new interaction
picture Hamiltonian
\[
\bar{H}_{\mathrm{int}}(t)= \lambda\,\nu \left(1-i\eta (e^{i\nu
t}\, a^\dagger + e^{-i\nu t}\, a)\right) e^{-i\delta
t}\,\sigma_- +\, h.c.
\]
--- which is once again time-dependent like the ion trap
Hamiltonian~{(\ref{initialham})} ---
and, then, retaining only those terms in
$\bar{H}_{\mathrm{int}}(t)$ which are `slowly rotating'; all the other terms,
often called `counter-rotating terms' (for historical reasons), are simply ignored.\\
In particular,
in correspondence to the three types of {\it resonance condition}
\begin{equation} \label{res}
\delta=\epsilon-\alpha\approx 0\, ,\ \ \ \delta +\nu\approx 0\, ,\ \
\ \delta -\nu\approx 0\, ,
\end{equation}
one obtains, respectively, the following three types of effective
interaction picture Hamiltonian:
\begin{eqnarray}
\heffo \spa & = & \spa \lambda\,\nu \left(\sigma_- +\,
\sigma_+\right),\hspace{2.32cm} (\,\mbox{\small $\delta\approx\,
0$}\,)
\\
\heffm \spa & = & \spa i\,\lambda\,\nu\,\eta
\left(a^\dagger\,\sigma_+ -\, a\,\sigma_-\right),\hspace{1cm}
(\,\mbox{\small $\delta\approx -\nu$}\,)
\\
\heffp \spa & = & \spa i\,\lambda\,\nu\,\eta \left(a\,\sigma_+ -\,
a^\dagger\,\sigma_-\right).\hspace{1.1cm} (\,\mbox{\small
$\delta\approx +\nu$}\,)
\end{eqnarray}
These effective Hamiltonians, in correspondence to the respective
resonances, commute with the reference Hamiltonian $\bar{H}_0$.
This is due to the fact that the resonances~{(\ref{res})} are
associated with the appearance of {\it degeneracies}
in the spectrum of the reference Hamiltonian,
and the degenerate eigenspaces of $\bar{H}_0$ are invariant subspaces for
the effective Hamiltonians $\heffo$, $\heffm$, $\heffp$,
respectively\footnote{In correspondence to the
resonances~{(\ref{res})}, the only non-degenerate eigenspaces of
$\bar{H}_0$ are given by
$\mathrm{span}(|0\rangle\otimes|+\rangle$), for $\delta=-\nu$, and
$\mathrm{span}(|0\rangle\otimes|-\rangle$), for $\delta=\nu$, that are invariant
subspaces for $\heffm$ and $\heffp$, respectively. All the other
eigenspaces are doubly degenerate.}
(in fact, it turns out that the spectrum of $\bar{H}_0$ is
degenerate if and only if the condition $|\delta|=m\,\nu\, ,\
m=0,1,2,\ldots\;$, holds). As a consequence, the evolution operators
generated by the effective Hamiltonians $\heffo$, $\heffm$, $\heffp$ can be
explicitly determined.\\
Notice moreover that, in particular, $\heffp$
--- up to a unitary transformation
\[
e^{i\pi\,\hat{n}/2}\; \heffp\; e^{-i\pi\,\hat{n}/2}
= \lambda\,\nu\,\eta \left(a\,\sigma_+ +\, a^\dagger\,\sigma_-\right)
\]
--- has the same form of the typical interaction term
of the Jaynes-Cummings Hamiltonian (see ref.~{\cite{Jaynes}}).

At this point, it is quite natural to address the following questions:
\begin{enumerate}

\item Is it {\it really} necessary, after having obtained the time-independent Hamiltonian
$\bar{H}$, to switch to a further interaction picture in order to
achieve a `good approximate Hamiltonian' --- precisely, an
effective time-independent interaction picture Hamiltonian such
that the associated evolution operator can be explicitly
determined --- as it is done for applying the RWA?

\item Moreover, does the RWA {\it really} allow to obtain a correct first order approximation
(with respect to the perturbative parameter $\lambda$) of the evolution operator?

\item Finally, is it possible to cast in a unique theoretical framework the perturbative
treatment of time-independent and time-dependent Hamiltonians in such a
way that, for instance, one can study both the time-dependent Hamiltonian
$\ho+H_\updown(t)$ and the time-independent Hamiltonian
$\bar{H}_0+\bar{H}_\updown$ using essentially the same approach (and, hopefully, finding
comparable results)?

\end{enumerate}
As it will be seen later on, the answers to these questions (in the
respective order) are the following:
\begin{enumerate}

\item No, it is not necessary. One can apply a time-independent
perturbative approach that allows to obtain approximate expressions of
the evolution operator in the remarkable form of a one-parameter group
of unitary transformations (see sects.~{\ref{independent}} and~{\ref{applications}}).

\item No, the RWA does not provide, already at the first
perturbative order, the correct expression of the evolution
operator associated with the Hamiltonian $\bar{H}$;\footnote{It
should be clear that this statement concerns the evolution
operator itself and {\it does not} exclude the possibility that
the behavior of certain experimentally observable quantities, with
specific initial conditions of the system, may be rather well
predicted using the RWA; see the discussion in
sect.~{\ref{conclusions}}.} it allows only to reproduce the {\it
qualitative behavior} of the correct first order expression (see
sect.~{\ref{applications}}).

\item Yes. As it will be shown in the subsequent sections, one
can develop a suitable perturbative approach that allows to treat
on the same footing both time-independent and time-dependent
Hamiltonians (see sect.~{\ref{generalcase}}). For instance, one
can apply this approach to the time-dependent Schr\"odinger
picture Hamiltonian $\ho+H_\updown(t)$ (or $\ho + H_\sapprox(t)$)
and to the time-independent interaction picture Hamiltonian
$\bar{H}_0+\bar{H}_\updown$ obtaining the same result, i.e.\ the
same perturbative expansion of the (Schr\"odinger picture)
evolution operator (see sect.~{\ref{applications}}).

\end{enumerate}


\section{Basic assumptions and strategy}
\label{basic}

Let us consider a time-dependent perturbed Hamiltonian $\hlt$,
namely a selfadjoint linear operator of the
form
\begin{equation} \label{hamiltonian}
\hlt=\ho(t)+\hilt\, ,
\end{equation}
where $\ho(t)$ is a selfadjoint (and, in general, time-dependent)
operator --- the `unperturbed component' --- and $\hilt$ is a
time-dependent perturbation; precisely, we will assume that
$\lambda\mapsto\hilt$ is (for the perturbative parameter $\lambda$ in a certain
neighborhood of zero and for any $t$) a real analytic,
selfajoint
operator-valued function, with $\hi(0;t)=0$.
A real analytic function can be extended to a domain in the complex
plane. Keeping this fact in mind, we will specify that a given property
{\it holds for $\lambda$ real}. For instance, the analytic function $\lambda\mapsto\hilt$
will take values in the selfadjoint operators for $\lambda$ real only.

Let $U(\lambda;t,t_0)$ be the evolution operator associated with
$\hlt$, with initial time $t_0$:
\begin{equation}
i\,\dot{U}(\lambda;t,t_0)=\hlt\,U(\lambda;t,t_0)\, , \ \ \
U(\lambda ;t_0,t_0)=\mathrm{Id}\, ,
\end{equation}
where the dot denotes the time derivative and we have set $\hbar=1$. Then, we have that
\begin{equation}
U(\lambda;t,t_0)= \uo(t,t_0)\, \rotev(\lambda;t,t_0)\, ,
\end{equation}
where $\uo(t,t_0)$ and
$\rotev(\lambda;t,t_0)$
are respectively the evolution operator associated with the unperturbed
component $\ho(t)$ (evolution operator which, if the unperturbed Hamiltonian
is time-independent, $\ho(t)\equiv\ho$, is obviously given by
$e^{-i\ho (t-t_0)}$)
and the evolution operator associated with the
interaction picture Hamiltonian
\begin{equation}
\rotham(\lambda;t,t_0) := \uo(t_0,t)\, \hilt \, \uo(t,t_0)\, .
\end{equation}
Let us notice explicitly that, since $\rotham(0;t,t_0)=0$, we have:
\begin{equation} \label{condi}
T(0;t,t_0)=\mathrm{Id}\, .
\end{equation}

We will suppose that the unperturbed evolution $\uo(t,t_0)$ is
explicitly known. Then the problem is to determine perturbative
expressions of $\rotev(\lambda;t,t_0)$. To this aim, the central
point of the paper is the assume for $\rotev(\lambda;t,t_0)$ the
following general decomposition:
\begin{equation} \label{gen1}
\!\!\rotev(\lambda;t,t_0) = \exp\left(-i\,Z\lto\right)\,
\exp\!\left(\!\!-i\!\int_{t_0}^t\!C(\lambda;\ttt,t_0)\,\de\ttt\!\right)
\exp\left(i\,Z(\lambda;t_0,t_0)\right),
\end{equation}
where $(\lambda ;t)\mapsto Z\lto$, $(\lambda ;t)\mapsto C\lto$ are
operator-valued functions which depend analytically on the
perturbative parameter $\lambda$;
in agreement with condition~{(\ref{condi})}, we set:
\begin{equation}
Z(0;t,t_0)=0\, ,\ \ \ C(0;t,t_0)=0\, ,\ \ \ \ \forall t\, .
\end{equation}
We stress that the presence of the term $\exp(i\,Z(\lambda;
t_0,t_0))$ in formula~{(\ref{gen1})} ensures that
$T(\lambda;t_0,t_0)=\mathrm{Id}$, allowing the possibility that
$Z(\lambda;t,t_0)\neq 0$ for $t=t_0$.

It will be seen that decomposition~{(\ref{gen1})} has a wide range
of solutions and that a possible choice for fixing a certain class
of solutions is given by imposing the condition $C\lto=\cl$, i.e.
assuming that the function $(\lambda ;t)\mapsto C\lto$ does not
depend on time. This decomposition includes, as particular cases,
two decompositions of the evolution operator that have been
considered in the literature:
\begin{itemize}
\item the decomposition that is obtained setting
$
Z\lto=0,\ \forall t,
$
in formula~{(\ref{gen1})}, decomposition
which is at the root of the {\it Magnus expansion} of the
evolution operator~{\cite{Magnus}};
\item the classical {\it Floquet decomposition} that holds in the
case where the interaction picture Hamiltonian depends
periodically on time (say with period $\period$) --- decomposition
which is obtained setting $ C\lto\equiv\cl, \ \
Z(\lambda;t_0,t_0)=0, $ and assuming that $(\lambda,t)\mapsto
Z(\lambda;t,t_0)$ is periodic with respect to time with period
$\period$ --- and that is at the root of the {\it Floquet-Magnus
expansion} of the evolution operator~{\cite{Casas}}.
\end{itemize}

From this point onwards, for notational convenience, we will fix
$t_0=0$. Then, decomposition~{(\ref{gen1})} can be rewritten as
\begin{equation} \label{gen2}
\rotev\lt =
\exp\left(-i\zlt\right)\,\exp\!\left(-i\cltint\right)\,\exp\left(i\zl\right),
\end{equation}
where:
$
\rotev\lt\equiv \rotev(\lambda;t,0),\ \zlt\equiv Z(\lambda;t,0),\
\zl\equiv Z(\lambda; 0), \ \clt\equiv C(\lambda;t,0).
$
Let us now proceed to obtain perturbative expansions of
$\clt$ and $\zlt$. To this aim, if we require the interaction picture
evolution operator to satisfy the Schr\"odinger equation, we get:
\begin{eqnarray}
\rothamlt\,\rotev\lt
\spa & = & \spa i\,\dot\rotev\lt
\nonumber \\
& = & \spa
e^{-i\zlt}\!\!\int_{0}^{1}\!\!\!\left(e^{is\zlt}\,\dot{Z}\lt\,e^{-is\zlt}\right)\!
\de s\ e^{-i\cltint} e^{i\zl} + \nonumber \\
& & \spa\spa\spa\spa\spa\spa\spa\spa\spa\spa +\
e^{-i\zlt}\!\!\int_{0}^{1}\!\!\!\left(e^{-is\cltint}\,\clt\,
e^{is\cltint}\right)\!
\de s\ e^{-i\cltint} e^{i\zl},
\label{preceding}
\end{eqnarray}
where we have used the remarkable formula (see, for instance,
ref.~\cite{Wilcox})
\begin{equation}
\der\, e^F = e^F \int_{0}^{1} \!\left(e^{-sF}\, \dot{F} \,
e^{sF}\right)\! \de s = \int_{0}^{1}\! \left(e^{sF}\, \dot{F}\,
e^{-sF}\right)\!\de s\  e^F ,
\end{equation}
which extends to an operator-valued function $t\mapsto F(t)$ the
formula for the derivative of the exponential of an
ordinary function. Next, let us apply to each member of
eq.~{(\ref{preceding})} the operator $e^{i\zlt}$ on the left
and the operator $e^{-i\zl}e^{i\cltint}$ on the right:
\begin{equation} \label{prestart}
\Ad_{\exp(i\zlt)}\,\rothamlt = \int_0^1 \!
\left(\Ad_{\exp(is\zlt)}\,\dot{Z}\lt +
\Ad_{\exp\left(-is\cltint\right)}\,\clt\right)\!\de s \, ,
\end{equation}
where we recall that, given linear operators $\X,Y$, with $\X$
invertible,
$
\Ad_{\X}\,Y:=\X\, Y\,\X^{-1}.
$
Then, since $\X$ is of the form $e^X$, we can use the well known
relation
\begin{equation} \label{form}
\Ad_{\exp(X)}\, Y=\exp(\ad_X)\, Y =\sum_{k=0}^\infty
\frac{1}{k!}\,\ad_X^k\, Y \, ,
\end{equation}
with $\ad_X^k$ denoting the $k$-th power ($\ad_X^0\equiv\mathrm{Id}$)
of the superoperator $\ad_X$ defined by
$
\ad_X\,Y:=[X,Y].
$
Eventually, applying formula~{(\ref{form})} to eq.~{(\ref{prestart})}
and performing the integrals, we obtain:
\begin{equation} \label{eqcompl}
\sum_{k=0}^\infty \frac{i^k}{k!}\,\ad_{\zlt}^k\, \rothamlt =
\sum_{k=0}^\infty \frac{i^k}{(k+1)!}\,\ad_{\zlt}^k\, \dot{Z}\lt +
\sum_{k=0}^\infty \frac{(-i)^k}{(k+1)!}\,\ad_{\cltint}^k\,\clt \,
.
\end{equation}
This equation will be the starting point for the determination of
the operator-valued functions $(\lambda,t)\mapsto\zlt$ and
$(\lambda,t) \mapsto\clt$ at each perturbative order in $\lambda$,
task that will be pursued systematically in the next sections.


\section{The time-independent case}
\label{independent}

We will first consider the most important special case: the case
where the Hamiltonian~{(\ref{hamiltonian})} does not depend on
time. We have at least three good reasons to single out this case
and to study it at the beginning:
\begin{itemize}
\item to show that in this case --- `the simplest one' --- our
perturbative decomposition is far from being trivial;

\item to highlight the link between our approach and standard perturbation theory for
linear operators, a link that on the other hand is completely
missing for Dyson and Magnus expansions;

\item to show how this special case can be regarded as a
natural starting point for extending the approach to the general
time-dependent case (`induction versus deduction').
\end{itemize}
As it will be clear soon, it will be now convenient to set
\begin{equation}
\zz\lt:=\uot\,\zlt\,\uot^\dagger,\ \ \
\zz(\lambda)\equiv\zz(\lambda;0)=\zl \, ,
\end{equation}
and re-express eq.~{(\ref{eqcompl})} in terms of the transformed
operator $\zz\lt$. To this aim, let us first notice that
\begin{equation} \label{derivative}
\dot{Z}\lt=\Ad_{\uot^\dagger}\left(\dot{\zz}\lt-i\,\ad_{\zz\lt}\,\ho(t)\right).
\end{equation}
Besides, given linear operators $\X$, $X$ and $Y$, with $\X$
invertible, one can show easily that
\begin{equation} \label{relad}
\ad_{\Ad_{\X} X}^k\,\Ad_{\X}\, Y=\Ad_{\X}\,\ad_X^k\, Y,\ \ \
k=0,1,2,\ldots\ .
\end{equation}
Then, since $\zlt=\Ad_{\uot^\dagger}\,\zz\lt$,
$\rothamlt=\Ad_{\uot^\dagger}\,\hilt$
and relation~{(\ref{derivative})} holds, using formula~{(\ref{relad})},
from eq.~{(\ref{eqcompl})} we obtain:
\begin{eqnarray}
\!\!\! \Ad_{\uot^\dagger}\sum_{k=0}^\infty
\frac{i^k}{k!}\,\ad_{\zz\lt}^k\,\hilt \spa & = & \spa
\Ad_{\uot^\dagger}\Big(\sum_{k=0}^\infty
\frac{i^k}{(k+1)!}\,\ad_{\zz\lt}^k\,\dot{\zz}\lt \nonumber\\
& - & \spa \sum_{k=1}^\infty \frac{i^k}{k!}\,\ad_{\zz\lt}^k\,\ho(t)\Big)
+ \sum_{k=0}^\infty
\frac{(-i)^k}{(k+1)!}\,\ad_{\cltint}^k\,\clt. \nonumber
\end{eqnarray}
Next, applying the superoperator $\Ad_{\uot}$ to each
member of this equation and rearranging the terms, we get
\begin{eqnarray}
\spa  \sum_{k=1}^\infty
\frac{i^k}{k!}\,\ad_{\zz\lt}^k\!\left(\ho(t)+\hilt\right) +\hilt \spa
& = &
\spa \Ad_{\uot}
\sum_{k=0}^\infty
\frac{(-i)^k}{(k+1)!}\,\ad_{\cltint}^k\,\clt \nonumber\\
& + & \spa \sum_{k=0}^\infty
\frac{i^k}{(k+1)!}\,\ad_{\zz\lt}^k\,\dot{\zz}\lt \, . \label{gen3}
\end{eqnarray}
This equation is, in general, harder to solve than
eq.~{(\ref{eqcompl})}; but, in the {\it time-independent case}, we
have that $\ho(t)\equiv\ho$, $\hilt\equiv\hil$, and it is natural
to assume:
\begin{equation} \label{tindcond}
\clt=\cl\, ,\ \ \ \zz\lt =\zz (\lambda ;0)=Z(\lambda ;0)\equiv\zl
\, .
\end{equation}
Then, eq.~{(\ref{gen3})} can be recast in a much simpler form:
\begin{equation} \label{cost}
\sum_{k=1}^\infty
\frac{i^k}{k!}\,\ad_{\zl}^n\!\left(\ho+\hil\right)+\hil = e^{-i\ho
t}\,\cl\, e^{i\ho t}.
\end{equation}
Now, observe that the first member of this equation does not
depend on time, hence the function $t\mapsto e^{-i\ho t}\,\cl\,
e^{i\ho t}$ must be constant. It follows that, if we want
eq.~{(\ref{cost})} to be consistent, we have to assume also that $
[\cl,\ho]=0, $ i.e. that $\cl$ {\it is a constant of the motion
for the unperturbed evolution generated by} $\ho$. Eventually, we
find:
\begin{equation} \label{decomposition}
\sum_{k=1}^\infty
\frac{i^k}{k!}\,\ad_{\zl}^k\!\left(\ho+\hil\right) +\hil =
C(\lambda) \, .
\end{equation}
At this point, we are ready to obtain perturbative expansions of
the operators $\cl$ and $\zl$ (hence, of the interaction picture
evolution operator $\rotev(\lambda ;t)$). We will suppose that the
unperturbed Hamiltonian $\ho$ has a pure point spectrum, while the
case where this hypothesis is not satisfied is a particular case
of the general treatment developed in sect.~{\ref{generalcase}}.
We will denote by $E_1,E_2,\ldots$ the (possibly degenerate)
eigenvalues of $\ho$ and by $P_1,P_2,\ldots$ the associated
eigenprojectors. Since the functions $\lambda\mapsto\hil$,
$\lambda\mapsto C(\lambda)$ and $\lambda\mapsto Z(\lambda)$ are
analytic and $\hi(0)=C(0)=Z(0)=0$, we can write:
\begin{equation} \label{power}
\hil = \sum_{n=1}^\infty \lambda^n\, H_n \, ,\ \ \
C(\lambda)=\sum_{n=1}^{\infty}\lambda^n\, C_n \, ,\ \ \
Z(\lambda)=\sum_{n=1}^{\infty}\lambda^n\, Z_n \, .
\end{equation}
Now, in order to determine the operators
$\{C_n\}_{n\in\mathbb{N}}$ and $\{Z_n\}_{n\in\mathbb{N}}$, let us
substitute the power expansions~{(\ref{power})} in
eq.~{(\ref{decomposition})}; in correspondence to the various
orders in the perturbative parameter $\lambda$, we get the
following set of conditions:
\begin{eqnarray}
C_1 -i\left[Z_1,\ho\right]-H_1=0 \!\! & , & \!\!
\left[C_1,\ho\right]=0 \label{first}
\\
\spa\spa\spa
C_2-i\left[Z_2,\ho\right]+\frac{1}{2}\left[Z_1,\left[Z_1,\ho\right]\right]
-i\left[Z_1,H_1\right]-H_2=0 \!\! & , & \!\!
\left[C_2,\ho\right]=0 \label{second}
\\
& \vdots & \nonumber
\end{eqnarray}
where we have taken into account the additional constraint
$[C(\lambda),\ho]=0$. This infinite set of equations can be solved
recursively and the solution --- as it should be expected (we will
clarify this point soon) --- is not unique.
The first equation, together with the first constraint, determines
$Z_1$ up to an operator commuting with $\ho$ and $C_1$ uniquely.
Indeed, since
\begin{equation}
[C_1,\ho]=0\   \Rightarrow\ C_1=\sum_{m} P_m\,C_1\,P_m \, ,\ \ \ \
\mbox{and}\ \ \ \ [Z_1,\ho]=\sum_{j\neq l}
\left(E_l-E_j\right)P_j\,Z_1\,P_l \, ,
\end{equation}
we conclude that
\begin{equation} \label{formulac}
C_1 = \sum_{m} P_m\,H_1\, P_m \, ,\ \ \ \ \mbox{and}\ \ \ \ Z_1=
\sum_{m} P_m\,Z_1\, P_m + i\sum_{j\neq l}
\left(E_l-E_j\right)^{-1} P_j\,H_1\,P_l \, .
\end{equation}
This last equation admits a {\it minimal solution} which is
obtained by imposing a further condition, namely:
$
P_m\, Z_1\, P_m =0, \ \ \ m=1,2,\ldots\ .
$

For $n>1$, we will adopt an analogous reasoning. Given an
operator $X$, let us set
\begin{equation} \label{ef}
\GG_{n}(X; Z_1,\ldots ,Z_{n}) := \sum_{m=1}^{n}\, \frac{i^m}{m!}
\!\!\! \sum_{\ \ \ k_1+\cdots +k_m=n} \!\! \ad_{Z_{k_1}} \cdots\
\ad_{Z_{k_m}}\, X \, ,
\end{equation}
with $n\ge 1$. Then, for $n\ge 2$, we can define the operator function
\begin{equation} \label{gi}
\spa\GGG_n(\ho,\ldots ,\hn ;Z_1,\ldots ,Z_{n-1})  :=
\sum_{m=0}^{n-1} \GG_{n-m}(\hm;Z_1,\ldots ,Z_{n-m}) - i[Z_n,\ho]+
H_n \, .
\end{equation}
At this point, one can show that the sequence of equations
generated by formula~{(\ref{decomposition})} is given by
\begin{eqnarray}
C_1 -i\left[Z_1,\ho\right]=H_1 \!\! & , & \!\!
\left[C_1,\ho\right]=0
\nonumber\\
& \vdots & \nonumber\\
\spa \spa \spa C_n - i\left[Z_n,\ho \right] = \GGG_n (\ho,\ldots
,\hn; Z_1,\ldots, Z_{n-1}) \!\! & , &  \!\! \left[C_n,\ho\right]=0
\ \ \ \ n\ge 2 \label{general}
\\
& \vdots & \nonumber
\end{eqnarray}
In order to write the general solution of this sequence of
equations, it will be convenient to introduce a shorthand
notation; given a linear operator $X$, we set:
\begin{equation}
\infra{X} := \sum_m P_m\, X\,P_m \, ,\ \ \ \extra{X} :=
X-\infra{X}=\sum_{j\neq l} P_j\, X\, P_l \, ,
\end{equation}
\begin{equation}
\Extra{X} := i \sum_{j\neq l}(E_l-E_j)^{-1} P_j\, X \, P_l \, .
\end{equation}
Now,
assume that the first $n$ equations have been solved. Then, the
operator function $\GGG_{n+1}(\ho,\ldots ,H_{n+1}; Z_1,\ldots,Z_n)$ is
known explicitly and hence
\begin{equation} \label{forc}
C_{n+1}= \infra{\GGG_{n+1}(\ho,\ldots ,H_{n+1};
Z_1,\ldots,Z_n)},
\end{equation}
\begin{equation} \label{forz}
\left[Z_{n+1},\ho\right]= i\extra{\GGG_{n+1}(\ho,\ldots
,H_{n+1}; Z_1,\ldots,Z_n)}.
\end{equation}
Again, this last equation determines $Z_{n+1}$ up to
an arbitrary operator $\infra{Z_{n+1}}$
commuting with $\ho$; in fact, we have:
\begin{equation}
Z_{n+1}=\infra{Z_{n+1}}+
\Extra{\GGG_{n+1}(\ho,\ldots,H_{n+1};Z_1,\ldots,Z_n)}.
\end{equation}
We stress that, in general, the choice of a particular solution
for $Z_{n+1}$ will also influence the form of
$C_{n+2},Z_{n+2},\ldots\ $. Thus, we conclude that the sequence of
equations defined above admits infinite solutions (even in the
case where $\ho$ has a non-degenerate spectrum). However, there is
a unique {\it minimal solution}
$\{\minc_n,\minz_n\}_{n\in\mathbb{N}}$ which fulfills the
following additional condition:
\begin{equation} \label{minsol}
\infra{\minz_n}=0,\ \ \ \ n=1,2,\ldots\ .
\end{equation}

To clarify the link of our approach with standard
perturbation theory for linear operators, let us recall a few
facts (see~{\cite{Kato-pert}}~{\cite{Reed}}). It is possible to show
that, under certain technical conditions, there exist positive
constants $r_1,r_2,\ldots$ and a simply connected neighborhood
$\mathcal{I}$ of zero in $\mathbb{C}$ such that, for any
$\lambda\in\mathcal{I}$ and $m=1,2,\ldots\;$, one has that:

\begin{description}
\item[1)]\hspace{0.22cm}
the following
contour integral on the complex plane
\begin{equation} \label{proj}
P_m(\lambda)=\frac{1}{2\pi i}\oint_{\Gamma_m}\!\!\!\! dz \ \,
\left(z-\hl\right)^{-1}
\end{equation}
--- where $\Gamma_m$ is the anticlockwise oriented circle $[0,2\pi]\ni\theta\mapsto
E_m + r_m\, e^{i\theta}$ around the eigenvalue $E_m$
--- defines a projection ($P_m(\lambda)^2=P_m(\lambda)$), which is an
orthogonal projection for $\lambda\in\mathcal{I}\cap\mathbb{R}$,
with $P_m(0)=P_m$, and
$\mathcal{I}\ni\lambda\mapsto P_m(\lambda)$ is an analytic
operator-valued function;
\item[2)]\hspace{0.25cm}
the range of the projection $P_m(\lambda)$ is an invariant
subspace for $\hl$ (but, if the range of $P_m$ is not
1-dimensional, in general not an eigenspace), hence
\begin{equation} \label{inv}
\hl\, P_m(\lambda) = P_m(\lambda)\, \hl\, P_m(\lambda)\, ;
\end{equation}
\item[3)]\hspace{0.23cm}
there exists a (non-unique) analytic family $\lambda\mapsto
W(\lambda)$ of invertible operators such that
\begin{equation} \label{trasf}
P_m = W(\lambda)^{-1} P_m(\lambda) \, W(\lambda) \, ,\ \ \
W(0)=\mathrm{Id}
\end{equation}
--- with $W(\lambda)$ unitary for $\lambda$ real ---
which is solution of a Cauchy problem of the type
$
i\,W^\prime(\lambda)=J(\lambda)\,W(\lambda),\
W(0)=\mathrm{Id},
$
where the apex denotes the derivative with respect to the
perturbative parameter and $\lambda\mapsto J(\lambda)$ is any
analytic family of operators --- selfadjoint for $\lambda$ real
--- such that
\begin{equation}
\sum_{l\neq m} P_l(\lambda)J(\lambda) P_m(\lambda) =i\sum_m
{P_m\hspace{-2.1mm}}^\prime\ (\lambda)\,P_m(\lambda)\ \ \ \ \ \
(*\ P_m(\lambda)\, {P_m\hspace{-2.1mm}}^\prime\
(\lambda)\,P_m(\lambda)=0\ *).
\end{equation}
\end{description}
In standard (Rayleigh-Schr\"odinger-Kato) perturbation theory, one
can obtain the perturbative corrections to unperturbed eigenvalues
and eigenvectors exploiting  formula~{(\ref{proj})} and a suitable
expansion of the {\it resolvent operator} $(z-\hl)^{-1}$ (see, for
instance, ref.~{\cite{Messiah}}). In order to recover our previous
results, we can use, instead, properties~{{\bf 2)} and {\bf 3)}
(compare with~\cite{Aniello1}~\cite{Aniello3}). Indeed, let us
define the operator
\begin{equation}
\hb:=W(\lambda)^{-1} \hl\, W(\lambda)\, ,
\end{equation}
which, for $\lambda$ real, is unitarily
equivalent to $\hl$. Using relations~{(\ref{inv})} and
(\ref{trasf}), we find
\begin{equation}
\hb\, P_m = W(\lambda)^{-1} \hl\, P_m(\lambda)\,
W(\lambda)=  W(\lambda)^{-1} P_m(\lambda)\,\hl\,P_m(\lambda)\,
W(\lambda)
\end{equation}
and hence:
$
\hb\, P_m= P_m \,\hb\, P_m, \ \ \ m=1,2,\ldots\ .
$
It follows that
$
\left[\,\hb,\ho\right]=0
$
and then we obtain the following important relation:
\begin{equation}
\left[W(\lambda)^{-1} \hl\, W(\lambda) - \ho , \ho\right]=0 \, .
\end{equation}
Thus, if we set $ W(\lambda)^{-1} \hl\, W(\lambda) - \ho=\cl,\
W(\lambda)=\exp\!\left(-i\,\zl\right), $ and we apply
relation~{(\ref{form})}, we find precisely
eq.~{(\ref{decomposition})}.

Concluding our treatment of the time-independent case, it is worth
stressing that, due to conditions~{(\ref{tindcond})}, for the
overall evolution operator we have:
\begin{eqnarray}
U\lt \spa & = & \spa e^{-i\ho t}\, e^{-i\zlt}\, e^{-i\cl t}\,
e^{i\zl} \nonumber\\ \spa\spa\spa\spa
(*\, \zz\lt = e^{-i\ho t}\,\zlt\,e^{i\ho t}=\zl\,*)\; \ \ \ \ \ \ \ & = & \spa
e^{-i\zl}\, e^{-i\ho t}\, e^{-i\cl t}\, e^{i\zl} \nonumber\\
\label{arr} (*\, \left[\cl,\ho\right]=0\, *)\;\ \ \ \ \ \ \ & = & \spa  e^{-i\zl}\,
e^{-i\left(\ho +\cl\right)t}\, e^{i\zl},
\end{eqnarray}
or, more explicitly,
\begin{equation} \label{reduced}
U\lt  = e^{-i\zl}\,\sum_m
\exp\!\left(-i(E_m+\rrc_m(\lambda))t\right) P_m\;
e^{i\zl},
\end{equation}
where we have introduced the {\it reduced rank operators}
\begin{equation}
\rrc_m(\lambda):=C(\lambda)\,P_m=P_m\,C(\lambda)\,P_m\, , \ \ \
m=1,\ldots\ .
\end{equation}
Notice that the truncations at each perturbative order of the
expression~{(\ref{reduced})} retain the fundamental property of
forming one-parameter groups of unitary transformations. To make
this statement precise, let us introduce the following notation.
Given an analytic function $\lambda\mapsto
f(\lambda)=\sum_{n=0}^{\infty}\lambda^n\, f_n$ and fixed a perturbative
order $\enne$, we will set
\begin{equation}
f\enneind (\lambda):= \sum_{n=0}^{\enne} \lambda^n\, f_n\, ;
\end{equation}
moreover, given another analytic function $\lambda\mapsto
h(\lambda)$, we will set:
\begin{equation} \label{notation}
f(\lambda)\app h(\lambda)\ \ \
\stackrel{\mathrm{def}}{\Longleftrightarrow}\ \ \
f\enneind(\lambda)=h\enneind(\lambda)\, .
\end{equation}
Then, for the evolution operator associated with $\hlt$ we have that
\begin{equation} \label{truncation}
U (\lambda ;t) \app \exp\!\left(-i\,Z\enneind(\lambda)\right)\,
\exp\!\left(-i\,(\ho+ C\enneind(\lambda))\,t\right)\,\exp\!
\left(i\,Z\enneind(\lambda)\right),
\end{equation}
where $[C\enneind(\lambda),\ho]=0\,$. Thus, the $\enne$-th order
truncation of our perturbative decomposition of the evolution
operator, i.e.\ the r.h.s.\ of relation~{(\ref{truncation})}, is
indeed a one-parameter group of unitary operators.

We conclude this section observing that one can read the array of
eqs.~{(\ref{arr})} `proceeding from the bottom to the top',
namely, one may assume the decomposition
\[
U\lt= e^{-i\zl}\, e^{-i\left(\ho +\cl\right)t}\, e^{i\zl},\ \ \ \
[\cl,\ho]=0\, ,
\]
as a starting point in the case of a time-independent Hamiltonian
and induce from this case the general
decomposition~{(\ref{gen2})}. This is actually the path
that has led the author to find the results presented in the paper,
extending in a natural way the results previously found in the
time-independent case (compare with refs.~\cite{Aniello1,
Aniello2, AnielloH, Militello, Aniello3}).


\section{The general case}
\label{generalcase}

We will now consider eq.~{(\ref{eqcompl})} in its
full generality, equation which can be re-written as
\begin{equation} \label{equation}
\sum_{k=0}^\infty
\frac{i^k}{k!}\,\ad_{\zlt}^k\!\left(\rothamlt-\frac{1}{k+1}\,\dot{Z}\lt\right)
=  \cc\lt \, ,
\end{equation}
where
\begin{equation} \label{hermitian*}
\cc\lt := \clt+\sum_{k=1}^\infty
\frac{(-i)^k}{(k+1)!}\,\ad_{\cltint}^k\,\clt \, .
\end{equation}
The operator $\clt$ can be recovered from the operator $\cc\lt$ by
means of an order by order procedure. Thus, we can solve
eq.~{(\ref{equation})} for $\cc\lt$ up to a given perturbative
order and obtain the perturbative expansion of $\clt$ truncated at
the same order parallely. Indeed, if we substitute in
eq.~{(\ref{hermitian*})} the power expansions
$\clt=\sum_{n=1}^{\infty}\lambda^n\, C_n(t)$ and
$\cc\lt=\sum_{n=1}^{\infty}\lambda^n\, \cc_n(t)$, and we single
out the various perturbative orders, we conclude that the $n$-th
order, which on the l.h.s.\ is given simply by
$\lambda^n\,\cc_n(t)$, consists on the r.h.s.\ of $\lambda^n\,
C_n(t)$ plus a function of $C_1(t),\ldots,C_{n-1}(t)$ and
$\int_0^t C_1(\ttt)\ \de\ttt,\ldots ,\int_0^t C_{n-1}(\ttt)\
\de\ttt$. Thus, we can achieve an order by order solution. Indeed,
one finds out that $\clt$ can be obtained from $\cc\lt$ by the
following recursive procedure:
\begin{eqnarray}
C_1(t) \spa & = & \spa \cc_1(t) \, ,
\nonumber\\
& \vdots & \nonumber \\ \label{recpro*} C_n(t) \spa & = & \spa
\RRR_n \!\left(C_1(t),\ldots ,C_{n-1}(t); \int_0^t C_1(\ttt)\
\de\ttt,\ldots ,\int_0^t C_{n-1}(\ttt)\ \de\ttt \right) + \cc_n(t)
\,  ,\ \ \ \ n\ge 2 \, , \nonumber
\\
& \vdots &
\end{eqnarray}
where the operator functions $\RRR_n(\ldots)$ are defined as
follows. Given linear operators
\[
X,X_1,\ldots, X_{n}\ \ \ \mbox{and}\ \ \ Y_1,\ldots,Y_{n}\, ,
\]
let us set
\begin{equation}
\RR_{n}(X; Y_1,\ldots ,Y_{n}) := -\sum_{m=1}^{n}\, \frac{(-
i)^m}{(m+1)!} \!\!\!\!\!\! \sum_{\ \ \ k_1+\cdots +k_m=n}
\!\!\!\!\!\!\! \ad_{Y_{k_1}} \cdots\ \ad_{Y_{k_m}}\, X \, , \; \ \
n\ge 1 \, .
\end{equation}
Then, for $n\ge 2$, we can define the operator function
\begin{equation} \label{defRRR}
\RRR_n(X_1,\ldots ,X_{n-1};Y_1,\ldots ,Y_{n-1}) :=
\sum_{m=1}^{n-1} \RR_{n-m}(X_m;Y_1,\ldots ,Y_{n-m}) \, .
\end{equation}

Let us now investigate the perturbative solutions of
eq.~{(\ref{equation})}. Substituting the power expansions $
\rotham\lt = \sum_{n=1}^\infty \lambda^n\, \rotham_n(t),\
\cc\lt=\sum_{n=1}^{\infty}\lambda^n\, \cc_n(t),\   \zlt
=\sum_{n=1}^{\infty}\lambda^n\, Z_n(t), $ we obtain an infinite
set of coupled equations that allows to compute order by order the
operators $\{\cc_n(t)\}_{n\in\mathbb{N}}$,
$\{Z_n(t)\}_{n\in\mathbb{N}}$. In fact, for $n\ge 1$, let us set
\begin{equation}
\breve{\mathcal{G}}_{n}(X,Y; Z_1,\ldots ,Z_{n}) := \sum_{m=1}^{n}
\frac{i^m}{m!} \!\!\!\!\!\!\! \sum_{\ \ \ k_1+\cdots +k_m=n}
\!\!\!\!\!\!\!\!\! \ad_{Z_{k_1}} \cdots\
\ad_{Z_{k_m}}\!\!\left(X-\frac{Y}{m+1}\right)\!.
\end{equation}
Then we can define $\hG_n(\rotham_1(t),\ldots ,\rotham_n(t) ;
Z_1(t),\ldots ,Z_{n-1}(t); \dot{Z}_1(t),\ldots ,\dot{Z}_{n-1}(t))$
as
\begin{equation} \label{defhG}
\sum_{m=1}^{n-1}
\breve{\mathcal{G}}_{n-m}(\rotham_m(t),\dot{Z}_m(t);Z_1(t),\ldots
,Z_{n-m}(t))+\rotham_n(t)\, , \ \ \ n\ge 2 \, .
\end{equation}
With these notations, one can write the sequence of coupled
equations which gives a perturbative solution of eq.~{(\ref{equation})}
as follows:
\begin{eqnarray}
\dot{Z}_1(t) \spa & = & \spa \rotham_1(t)-\cc_1\lt \, ,
\nonumber\\
& \vdots & \nonumber \\
\dot{Z}_n(t) \spa & = & \spa \hG_n\! \left(\rotham_1(t),\ldots
,\rotham_n(t); Z_1(t),\ldots, Z_{n-1}(t); \dot{Z}_1(t),\ldots,
\dot{Z}_{n-1}(t)\right) \nonumber\\ \label{eqset} & - & \spa
\cc_n(t) \, ,\ \ \ \ \ n\ge 2 \, ,
\\
& \vdots & \nonumber
\end{eqnarray}
As in the time-independent case, this infinite set of equations
can be solved recursively. Moreover, by virtue of the recursive
procedure~{(\ref{recpro*})}, one can calculate order by order both
the operators $\{C_n(t)\}_{n\in\mathbb{N}}$ and
$\{Z_n(t)\}_{n\in\mathbb{N}}$. Indeed, integrating with respect to
time each equation in the sequence~{(\ref{eqset})} and combining
the new sequence of equations so obtained with the recursive
process~{(\ref{recpro*})}, we find
\begin{eqnarray}
\hspace{-4.7cm} Z_1(t) \spa & = &  \spa \int_0^t\!
\left(\rotham_1(\ttt)-\cc_1(\ttt)\right)\de \ttt +  Z_1  \, ,
\nonumber\\ \label{firstcouple}
\hspace{-4.7cm} C_1(t) \spa & = &
\spa \cc_1(t) \, ,
\end{eqnarray}
at the the first order, and
\begin{eqnarray}
Z_2(t) \spa & = & \spa
\int_0^t\!\left(\hG_2\big(\rotham_1(\ttt),\rotham_2(\ttt);
Z_1(\ttt);\dot{Z}_1(\ttt)\big)-\cc_2(\ttt)\right)\de\ttt + Z_2
\nonumber\\
& = & \spa
\int_0^t\!\left(i\,\ad_{Z_1(\ttt)}\!\left(\rotham_1(\ttt)-\frac{1}{2}\,
\dot{Z}_1(\ttt)\right)+\rotham_2(\ttt)-\cc_2(\ttt)\right)\de\ttt +
Z_2\, , \nonumber
\end{eqnarray}
\begin{eqnarray}
\hspace{-4.2cm} C_2(t) \spa & = & \spa
\RRR_2\!\left(C_1(t),\int_0^t C_1(\ttt)\de\ttt\right)+\cc_2(t)
\nonumber\\
\hspace{-4.2cm} & = & \spa \frac{i}{2}\,\ad_{\int_0^t
C_1(\ttt)\de\ttt} C_1(t)+\cc_2(t)\,,
\\
\hspace{-4.2cm} & \vdots & \nonumber
\end{eqnarray}
In general, at the $n$-th order, for $n\ge 2$, we have:
\begin{eqnarray}
\spa\spa\spa\spa Z_n(t) \spa & = & \spa \int_0^t \!\left(\hG_n
\big(\ldots ,\rotham_n(\ttt);\ldots, Z_{n-1}(\ttt); \ldots,
\dot{Z}_{n-1}(\ttt)\big)-\cc_n(\ttt)\right)\de\ttt + Z_n \, ,
\nonumber\\
\spa\spa\spa\spa C_n(t) \spa & = & \spa
\RRR_n\!\left(\ldots ,C_{n-1}(t); \ldots ,\int_0^t C_{n-1}(\ttt)\
\de\ttt \right)\! + \cc_n(t) \,  .
\end{eqnarray}
Here, differently from the time-independent case, at each
perturbative order we have a {\it couple} of equations. The
solution of the first couple~{(\ref{firstcouple})} is obtained by
choosing the arbitrary operator-valued function $t\mapsto\cc_1(t)$
and the arbitrary operator $Z_1$; similarly, the solution of the
$n$-th couple of equations, for $n\ge 2$, involves the previously
computed functions $t\mapsto C_1(t),\ldots,t\mapsto C_{n-1}(t),$
$t\mapsto Z_1(t),\ldots, t\mapsto Z_{n-1}(t)$ and requires the
choice of the arbitrary operator-valued function $t\mapsto
\cc_n(t)$ and of the arbitrary operator $Z_n$. {\it This choice
can be fitted according to computational convenience or adapted to
specific requirements}.\\
While the first point (computational
advantages) can be best appreciated by means of concrete examples
--- see sect.~\ref{applications} ---
the second one (adapting the solutions to some requirement) will
be illustrated considering two important issues. One of these ---
to characterize the solutions of our perturbative decomposition
given a certain class of interaction picture Hamiltonians --- will
be the subject of last part of the present section.
The other one is the following. In many
applications it is customary to study a physical system described
by an interaction picture Hamiltonian $\rothamlt$ using some
effective Hamiltonian $\eff\lt$ which is easier to treat, claiming
(usually, on the physical ground) that one can in such a way
achieve a satisfactory description of the system. Now, it is
natural to address the question:
\begin{quote}
{\em what is the relation, at each perturbative order, between the
evolution operator associated with the effective Hamiltonian and
the true evolution operator?}
\end{quote}
In order to establish a precise setting for this question, we will
first provide an interpretation of decomposition~{(\ref{gen2})}
which sheds light on its meaning. Notice that this decomposition
can be rewritten as
\begin{equation} \label{intpic}
\TC\lt=\TZ\lt^{-1}\, T\lt\,\TZ (\lambda;0),
\end{equation}
where
\begin{equation}
\TC\lt:=\exp\!\left(\!-i\cltint\right)\ \ \mbox{and}\ \ \;
\TZ\lt:=\exp\!\left(-i\,\zlt\right).
\end{equation}
Formula~{(\ref{intpic})} can be regarded as a passage to a further
`generalized interaction picture' performed on the Hamiltonian
$\rothamlt$. Indeed, let us observe that $\TZ\lt$ satisfies the
equation
\begin{eqnarray}
i\,\dot{T}_Z\lt \spa & = & \spa \zzzl\lt\,\TZ\lt \nonumber\\
& = & \spa \TZ\lt\,\zzzr\lt\, ,\ \ \ \ \mbox{with}\ \
\zzzr\lt=\TZ\lt^{-1}\,\zzzl\lt\,\TZ\lt\,,
\end{eqnarray}
--- where $\zzzl\lt,\, \zzzr\lt$ have the following explicit form:
\begin{equation} \label{intph}
\zzzl\lt= \sum_{k=0}^\infty
\frac{(-i)^k}{(k+1)!}\,\ad_{\zlt}^k\,\dot{Z}\lt\,,\ \ \ \
\zzzr\lt= \sum_{k=0}^\infty
\frac{i^k}{(k+1)!}\,\ad_{\zlt}^k\,\dot{Z}\lt
\end{equation}
--- or, equivalently, $i\left(\frac{\partial}{\partial
t}\TZ^{-1}\right)\!\lt=\zzzr\lt\,\TZ\lt^{-1}$. From this relation
and eq.~{(\ref{intpic})}, one finds that
\begin{equation} \label{sheq}
i\,\dot{T}_C\lt=\left(\TZ\lt^{-1}\,\rotham\lt\,\TZ\lt-\zzzr\lt\right)\TC\lt\,.
\end{equation}
Then,  by~{(\ref{intph})} and~{(\ref{sheq})}, one can conclude
that eq.~{(\ref{equation})} expresses precisely the fact that
$\cc\lt$ is the transformed Hamiltonian obtained by switching to
this `new interaction picture'; namely:
\[
\cc\lt=
\TZ\lt^{-1}\!\left(\rotham\lt-\zzzl\lt\right)\TZ\lt=\TZ\lt^{-1}\,\rotham\lt\,\TZ\lt-\zzzr\lt\,.
\]
It follows that
\begin{equation}
\TC\lt=\exp\!\left(\!-i\sum_{n=1}^\infty\lambda^n\int_0^t
C_n(\ttt)\,\de\ttt\right)
\end{equation}
is nothing but the Magnus expansion of the evolution operator
associated with the new interaction picture Hamiltonian
$\cc\lt$.\footnote{Again, we stress that we use here the term
`interaction picture' in a generalized sense.}\\
Hence, coming back to our initial question, it is now clear that,
setting $\cc\lt=\eff\lt$, decomposition~{(\ref{gen2})} provides
precisely an order by order comparison between the true evolution
operator $\rotev\lt$ and the effective one $\TC\lt$ (notice that
only the arbitrary constants $\{Z_n\}_{n\in\mathbb{N}}$ are still
to be fixed in order to determine the decomposition of
$\rotev\lt$). A remarkable example of such a comparison will be
given in sect.~{\ref{applications}}.

Let us now turn to the issue of characterizing a certain class of
realizations of our perturbative expansion. To classify the whole
range of possible solutions and their specific properties would be
obviously a problem far beyond the scope of the present paper, in
its full generality. Our aim is to study a wide but coherent class
of solutions that is particularly relevant for applications.\\ We
will first focus on the important class of solutions which is
determined by the condition: $ \cc_1(t)=\cc_1(0)\equiv
\cc_1,\ldots, \cc_n(t)=\cc_n(0)\equiv\cc_n,\ldots\ \ \forall t. $
This condition is equivalent to the following:
\begin{equation} \label{conc}
C_1(t)=C_1(0)\equiv  C_1,\ldots,  C_n(t)=C_n(0)\equiv C_n,\ldots\
\ \forall t \, .
\end{equation}
Moreover, if this condition holds, we have: $ C_1=\cc_1,\ldots,
C_n=\cc_n,\ldots\ $. Then the solution of the first equation ---
namely $Z_1(\{C_1,Z_1\};t)=\int_0^t \rotham_1(\ttt)\,\de\ttt -t\,
C_1+Z_1$ --- is fixed by the choice of the `arbitrary constants'
$C_1$ and $Z_1$. Inductively, the solution of the $n$-th equation
can be achieved by substituting the previously obtained solutions
$ t\mapsto Z_1(\{C_1,Z_1\},t),\ldots,t\mapsto
Z_{n-1}(\{C_k,Z_k\}_{k=1}^{n-1};t) $
--- that are fixed by the choice of the arbitrary
constants $C_1,\ldots,C_{n-1}$ and $Z_1,\ldots,Z_{n-1}$ --- of the
first $n-1$ equations in the formula
\begin{eqnarray}
& \vdots & \nonumber\\ \label{seq}
\!\! Z_n(\{C_k,Z_k\}_{k=1}^n;t)
\spa & = & \spa \int_0^t\! \hG_n\!\left(\ldots
;\ldots, Z_{n-1}(\{C_k,Z_k\}_{k=1}^{n-1};\ttt); \ldots ,
\dot{Z}_{n-1}(\{C_k,Z_k\}_{k=1}^{n-1};\ttt)\right)\! \de\ttt \nonumber\\
\label{zeta} & - & \spa t\, C_n+Z_n \, ,\ \ \ \ \ \ n\ge 2\, ,\\
& \vdots & \nonumber
\end{eqnarray}
which involves the $n$-th order arbitrary constants $C_n$ and
$Z_n$.\\
We will now consider a specific possible choice of the arbitrary
constants $\{C_n,Z_n\}_{n\in\mathbb{N}}$ up to a certain
perturbative order $\enne\in\mathbb{N}$. Suppose that the
following limits exist (the existence and the meaning of these
limits will be investigated presently):
\begin{eqnarray}
\!\!\infc_1\spa & := & \spa
\lim_{\tau\rightarrow\infty}\frac{1}{\tau} \int_0^\tau
\rotham_1(t)\ \de t \, ,
\nonumber \\
\!\!\infz_1 \spa & := & \spa \lim_{\tau\rightarrow\infty}
\left(-\frac{1}{\tau}\int_0^\tau \!\left(
\int_0^t\left(\rotham_1(\ttt)-\infc_1\right)\, \de\ttt\right)\,\de
t \right)
\nonumber \\
& = & \spa \lim_{\tau\rightarrow\infty}
\left(-\frac{1}{\tau}\int_0^\tau \!\left( \int_0^t\rotham_1(\ttt)\
\de\ttt\right)\de t +\frac{1}{2}\,\tau\,\infc_1 \right),
\nonumber \\
& \vdots &
\label{limits} \\
\!\!\infc_\enne  \spa & := & \spa \lim_{\tau\rightarrow\infty}
\left( \frac{1}{\tau} \int_0^\tau\! \hG_\enne\!\left(\ldots
;\ldots; \ldots , \dot{Z}_{\enne-1}(\{\infc_k,
\infz_k\}_{k=1}^{\enne -1}; t)\right) \de t \right),
\nonumber \\
\!\!\infz_\enne \spa & := & \spa \lim_{\tau\rightarrow\infty}
\bigg(\!\!\! -\frac{1}{\tau} \int_0^\tau\! \!\!\left(\int_0^t\!
\hG_\enne\!\left(\ldots ;\ldots; \ldots , \dot{Z}_{\enne
-1}(\{\infc_k, \infz_k\}_{k=1}^{\enne -1};\ttt)\right)
\de\ttt\right) \de t + \frac{1}{2}\,\tau\,\infc_\enne\!\!\bigg).
\nonumber
\end{eqnarray}
Then, one can set $\{C_n=\infc_n,\, Z_n=\infz_n\}_{n=1}^\enne$.
This particular choice has a remarkable property. Indeed, in the
time-independent case
--- $\ho(t)\equiv\ho$ and $\hilt\equiv\hi(\lambda)$ --- there is
a precise relation between the solution associated with the
arbitrary constants $\{\infc_n,\infz_n\}_{n\in\mathbb{N}}$
introduced here and the solutions obtained in
sect.~{\ref{independent}}. As we have done in that section, we
will assume that the unperturbed Hamiltonian $\ho$ has a pure
point spectrum. At this point, one can prove that:
\begin{enumerate}
\item
for any $\enne\in\mathbb{N}$, the limits~{(\ref{limits})} exist;

\item the minimal solution $\{\minc_n,\,\minz_n\}_{n\in\mathbb{N}}$
of the sequence of equations~{(\ref{general})}, i.e.\ the solution
obtained imposing condition~{(\ref{minsol})}, satisfies
\begin{equation}
\minc_n=\infc_n\,, \ \  \minz_n=\infz_n\,,\ \ \ \forall n\in\mathbb{N}\, ;
\end{equation}

\item the solution of the sequence of equations~{(\ref{seq})}
determined by the arbitrary constants
$\{\infc_n,\infz_n\}_{n\in\mathbb{N}}$ is such that
\begin{equation}
Z_n(\{\infc_k,\infz_k\}_{k=1}^{n-1};t)= e^{i\ho
t}\;\infz_n\;e^{-i\ho t},\ \ \ \forall n\in\mathbb{N}\, .
\end{equation}
\end{enumerate}
Notice that the operators $\{\infc_n,\infz_n\}_{n\in\mathbb{N}}$
can be calculated by formulae~{(\ref{limits})} where  the
eigenprojectors of $\ho$ --- that are involved in the formulae of
sect.~{\ref{independent}} --- do not appear.

We will now prove the existence of the limits~{(\ref{limits})}
(and provide a simple interpretation of their meaning) in two
important cases where the (Schr\"odinger picture) Hamiltonian $H$
of the quantum system is, in general, time-dependent:
\begin{itemize}

\item the case where the functions $t\mapsto\rotham_1(t),\ldots$ are periodic;

\item the case where the functions $t \mapsto\rotham_1(t),\ldots$ are operator-valued
trigonometric polynomials --- in the following, for the sake of
conciseness, just {\it trigonometric polynomials} --- namely functions
of time of the type
\begin{equation}
F(t)=\sum_{k=1}^m A_k\, e^{i\omega_k t}\, ,\ \ \ \omega_k\in\mathbb{R}\,,
\end{equation}
where $A_1,\ldots,A_m$ are operators in the Hilbert space of our
quantum system.

\end{itemize}
Periodic functions and trigonometric polynomials are examples of
{\it almost-periodic} (a.p.) functions (see~\cite{Bohr,
Amerio, Zaidman}).\footnote{The general definition
of a a.p.\ function is rather technical and will not be used in the
following, so we omit it here and we address the reader who may be
interested to the cited references. A typical example of a
$\mathbb{C}$-valued function which is a.p., but {\it not}
periodic, is provided by the function $t\mapsto (e^{i\varkappa_1
t}+\, e^{i\varkappa_2 t})$, where
$\varkappa_1,\varkappa_2\in\mathbb{R}$ are such that
$\varkappa_1/\varkappa_2$ is irrational.} This class of functions
has the relevant property of giving rise to a Fourier analysis
that generalizes the standard Fourier analysis associated with
periodic functions (a generalization
mainly due to the great mathematician Harald Bohr).\\
It is worth mentioning that, not only periodic functions, but also
trigonometric polynomials play a prominent role among a.p.\
functions, since any a.p.\ function can be suitably approximated
by a trigonometric polynomial
(consider, for instance, the standard Fourier expansion of a periodic function).\\
Let us recall a few properties that will be useful soon:
\begin{description}

\item[(a)]\hspace{0.9mm} for any a.p.\ function $F$ --- in particular, for any
periodic function or trigonometric polynomial ---
one can define the {\it mean value} $\mean{F}$ of $F$, namely it exists the
limit
\begin{equation}
\mean{F}:=\lim_{\tau\rightarrow\infty}\frac{1}{\tau} \int_0^\tau
F(t)\ \de t \, ;
\end{equation}
if $F$ is a periodic function,
with period $\period$, one can easily show that $\mean{F}$
coincides with the ordinary mean value of $F$ as a periodic
function, i.e.\ $\mean{F}=\frac{1}{\period} \int_0^{\period} F(t)\ \de
t$; if $F$ is a periodic function with period $\period$ (alternatively,
a trigonometric polynomial), then $F-\mean{F}$ is a zero-mean-valued periodic function
with period $\period$ (respectively, a zero-mean-valued trigonometric polynomial);

\item[(b)]\hspace{0.6mm} given a a.p.\ function $F$, if the {\it primitive}
\begin{equation} \label{primitiva}
\prim :\, \mathbb{R}\ni t\mapsto \int_0^t F(\ttt)\ \de\ttt
\end{equation}
is a a.p.\ function, then $\mean{F}=0$; in the case where, in particular,
$F$ is a periodic function (alternatively, a trigonometric polynomial),
the primitive $\prim$ is periodic (respectively, a trigonometric polynomial)
if and only if $\mean{F}=0$;

\item[(c)]\hspace{0.6mm} for any a.p.\ function $F$, one can define
the {\it essential primitive} of $F$,\footnote{This term refers
to the idea (typical of signal analysis) that
$\smallint\hspace{-2.1mm}\mbox{--}\hspace{-0.2mm} F$ contains the
essential spectral information about $F$; i.e.\ that $F$ can be
reconstructed from
$\smallint\hspace{-2.1mm}\mbox{--}\hspace{-0.2mm} F$ up to its
constant component $\mean{F}$ (`up to a constant offset').}
namely the function
\begin{equation}
\primitive F (t):=\smallint\hspace{-0.9mm}\big(F -
\mean{F}\big)(t) =\int_0^t F(\ttt)\ \de\ttt - t\,\mean{F}\, ;
\end{equation}
if $F$ is a periodic function with period $\period$, then its
essential primitive $\primitive F$ is periodic too, with the same
period; moreover, the essential primitive of a trigonometric
polynomial is still a trigonometric polynomial.

\end{description}

At this point, {\it we will assume that}
\begin{quote}
up to a certain order $\enne$, the coefficients
$
t\mapsto\rotham_1(t)\,,\ldots,\,t\mapsto\rotham_{\enne}(t)
$
of the perturbative expansion of the interaction picture Hamiltonian
$\lt\mapsto\rotham\lt$ are periodic functions sharing a common
period $\period$, or, alternatively, trigonometric polynomials.
\end{quote}
Then, we can show that the limits~{(\ref{limits})} exist. Indeed,
if the function $\rotham_1$ is a periodic function with period
$\period$ (alternatively, a trigonometric polynomial), then,
by property~{\bf (c)}, its
essential primitive $\primitive \rotham_1$ is a periodic function
with the same period (respectively, a trigonometric polynomial).
Notice, now, that
\begin{eqnarray}
\infc_1\spa & = & \spa \mean{\rotham_{1}}\, , \nonumber \\
\label{lim1}
\infz_1 \spa & = & \spa - \mmean{\primitive\rotham_1}
= -\mmean{\!\smallint\hspace{-0.9mm}\big(\rotham_1-
\mean{\rotham_1}\big)},
\end{eqnarray}
and
\begin{equation} \label{soluno}
Z_1(\{\infc_1,\infz_1\};t)=\primitive\rotham_1(t)
-\mmean{\primitive\rotham_1}\, ;
\end{equation}
hence, $t\mapsto Z_1(\{\infc_1,\infz_1\};t)$ is a
(zero-mean-valued) periodic function
with period $\period$ (respectively, a zero-mean-valued trigonometric polynomial).\\
Next, if $\enne\ge 2$, by this fact and our initial assumption,
we find that the function
\begin{equation}  \label{funcsec}
t\mapsto\hG_2\big(\rotham_1(t), \rotham_2(t)
;Z_1(\{\infc_1,\infz_1\};t) ; \dot{Z}_{1}(\{\infc_1, \infz_1\}; t)\big)
\end{equation}
is periodic with period $\period$ (respectively, a trigonometric
polynomial) and
\begin{equation}
\infc_2=\mmean{\hG_2\big(\ldots , \rotham_2\mbox{\small
$(\,\cdot\,)$} ;\ldots ; \dot{Z}_{1}(\{\infc_1, \infz_1\};
\mbox{\small $(\,\cdot\,)$})\big)};
\end{equation}
moreover, by property~{\bf (c)}, we have that the essential primitive of the
function~{(\ref{funcsec})} is periodic with period $\period$
(respectively, a trigonometric polynomial). As a consequence, the
limit $\infz_2$, being (up to a minus sign) the mean value of the essential primitive
of the function~{(\ref{funcsec})}, exists and the function
$t\mapsto Z_2(\{\infc_k,\infz_k\}_{k=1}^2;t)$ is periodic
with period $\period$ (respectively, a trigonometric polynomial) again. \\
For $\enne\ge 3$ --- re-iterating the same argument --- one finds out
that the limits~{(\ref{limits})} exist up to the order $\enne$. In fact,
one finds that, for $2\le n\le \enne$, the function
\begin{equation} \label{funcen}
t\mapsto\hG_n\big(\ldots, \rotham_n(t)
;\ldots ,Z_{n-1}(\{\infc_k,\infz_k\}_{k=1}^{n-1};t) ;\ldots , \dot{Z}_{n-1}(\{\infc_k,
\infz_k\}_{k=1}^{n-1}; t)\big)
\end{equation}
--- together with its essential primitive --- is periodic with period $\period$
(respectively, a trigonometric polynomial). Thus, we have:
\begin{eqnarray}
\spa\spa\spa\infc_n  \spa & = & \spa \mmean{\hG_n\big(\ldots,
\rotham_n\mbox{\small $(\,\cdot\,)$} ;\ldots ; \ldots ,
\dot{Z}_{n-1}(\{\infc_k, \infz_k\}_{k=1}^{n-1}; \mbox{\small
$(\,\cdot\,)$})\big)}\,, \nonumber \\ \label{limn}
\spa\spa\spa\infz_n \spa & = & \spa
-\mmean{\primitive\hspace{0.5mm}\hG_n\big(\ldots,\rotham_n\mbox{\small
$(\,\cdot\,)$} ;\ldots; \ldots , \dot{Z}_{n -1}(\{\infc_k,
\infz_k\}_{k=1}^{n -1};\mbox{\small $(\,\cdot\,)$})\big)}\,, \ \ \
2\le n\le\enne\,;
\end{eqnarray}
moreover, for any $2\le n\le\enne$, the $n$-th order solution
\begin{eqnarray}
\spa\spa\spa Z_n(\{\infc_k,\infz_k\}_{k=1}^n;t) \spa & = & \spa
\Big(\primitive\hspace{0.5mm}\hG_n\big(\ldots,\rotham_n\mbox{\small
$(\,\cdot\,)$} ;\ldots; \ldots , \dot{Z}_{n -1}(\{\infc_k,
\infz_k\}_{k=1}^{n -1};\mbox{\small $(\,\cdot\,)$})\big)\Big) (t) \nonumber
\\ \label{soln}
& - & \spa
\mmean{\primitive\hspace{0.5mm}\hG_n\big(\ldots,\rotham_n\mbox{\small
$(\,\cdot\,)$} ;\ldots; \ldots , \dot{Z}_{n -1}(\{\infc_k,
\infz_k\}_{k=1}^{n -1};\mbox{\small $(\,\cdot\,)$})\big)}
\end{eqnarray}
is periodic with period $\period$ (respectively, a trigonometric
polynomial).

In conclusion, we have shown that
\begin{quote}
{\em if the coefficients $t\mapsto\rotham_1(t)\, ,\ldots$ of the
perturbative expansion of the interaction picture Hamiltonian are
periodic functions sharing a common period $\period$}
(alternatively, trigonometric polynomials) {\em up to a certain
order $\enne\ge 1$, then the limits}~{(\ref{limits})} {\em exist,
and they are given by formulae}~{(\ref{lim1})} {\em
and}~{(\ref{limn})}{\em ; moreover, for any $n\le \enne$, the
$n$-th order solution $t\mapsto
Z_n(\{\infc_k,\infz_k\}_{k=1}^n;t)$, which is given by
formula}~{(\ref{soluno})} {\em or}~{(\ref{soln})}{\em , is a
periodic function with period $\period$} (respectively, a
trigonometric polynomial);
\end{quote}
in addition, observe that we have the further remarkable property
(recall the notation~{(\ref{notation})} introduced at
the end of sect.~{\ref{independent}}):
\begin{quote}
{\em if the functions
$t\mapsto\rotham_1(t)\,,\ldots,\,t\mapsto\rotham_\enne(t)$ are
periodic, sharing a common period $\period$, in the $\enne$-th
order approximation of the interaction picture evolution operator
\begin{equation}
\rotev (\lambda ;t) \app \exp\!\left(-i\,Z\enneind(\lambda
;t)\right)\, \exp\!\left(-i\,C\enneind(\lambda)\,t\right)\,\exp\!
\left(i\,Z\enneind(\lambda)\right),
\end{equation}
where
\begin{equation}
Z\enneind(\lambda ;t)= \sum_{n=1}^{\enne} \lambda^n\,
Z_n(\{\infc_k,\infz_k\}_{k=1}^n;t)\, ,
\end{equation}
\begin{equation}
C\enneind(\lambda)=\sum_{n=1}^{\enne} \lambda^n\, \infc_n\,,\ \ \
Z\enneind(\lambda)=Z\enneind(\lambda;0)=\sum_{n=1}^{\enne}
\lambda^n\, \infz_n\,,
\end{equation}
the function $\mathbb{R}\ni
t\mapsto\exp\!\left(-i\,Z\enneind(\lambda;t)\right)$ is
periodic with period $\period$.}
\end{quote}

Actually, one can easily prove (along the lines previously drawn)
that --- in the case where  the coefficients $t\mapsto
\rotham_1(t)\,,\ldots,\,t\mapsto\rotham_\enne(t)$ are periodic
functions sharing a common period $\period$ (alternatively,
trigonometric polynomials), if at each perturbative order
$n\le\enne$ one fixes the operator $Z_n$ {\it arbitrarily}, and
sets $C_1= \mean{\rotham_{1}}$ and, for $\enne\ge 2$,\footnote{One
can check recursively that the function $t\mapsto\hG_n\big(\ldots,
\rotham_n(t) ;\ldots ; \ldots , \dot{Z}_{n-1}(\{C_k,
Z_k\}_{k=1}^{n-1}; t)\big)$ (hence, its essential primitive), for
$2\le n\le\enne$, is periodic with period $\period$ (respectively,
a trigonometric polynomial).}
\[
C_n = \mmean{\hG_n\big(\ldots, \rotham_n\mbox{\small
$(\,\cdot\,)$} ;\ldots ; \ldots , \dot{Z}_{n-1}(\{C_k,
Z_k\}_{k=1}^{n-1}; \mbox{\small $(\,\cdot\,)$})\big)}\,,\ \ \ 2\le
n\le \enne\, ,
\]
so that
$Z_1(\{C_1,Z_1\};t)=\primitive\rotham_1(t)+Z_1$ and, for $\enne\ge
2$,
\begin{eqnarray}
\spa\spa\spa Z_n(\{C_k,Z_k\}_{k=1}^n;t) \spa & = & \spa
\Big(\primitive\hspace{0.5mm}\hG_n\big(\ldots,\rotham_n\mbox{\small
$(\,\cdot\,)$} ;\ldots
,Z_{n-1}(\{C_k,Z_k\}_{k=1}^{n-1};\mbox{\small
$(\,\cdot\,)$})
; \ldots
\big)\Big)(t) \nonumber
\\ \label{soln*}
& + & \spa Z_n\, , \ \ \ 2\le n\le\enne\, ,
\end{eqnarray}
--- {\it the preceding results remain true}; namely: the function
$t\mapsto Z_n(\{C_k,Z_k\}_{k=1}^n;t)$, for any $n\in\{1,\ldots
,\enne\}$, is periodic with period $\period$ (respectively, a
trigonometric polynomial). Thus, our perturbative expansion of the
evolution operator can indeed be regarded as a generalization of
the Floquet-Magnus expansion, which is recovered
--- in the case where the functions $t\mapsto
\rotham_1(t),\ldots,t\mapsto\rotham_\enne(t)$ are periodic,
sharing a common period --- by setting in particular:
$Z_1=0\,,\ldots, Z_\enne=0\,$.


\section{The ion trap Hamiltonian (revisited)}
\label{applications}

We will now re-consider, in the light of the theory developed in
sects.~{\ref{basic}}, \ref{independent} and~{\ref{generalcase}},
the quantum Hamiltonian describing a trapped two-level ion
interacting with a monochromatic laser field (in the Lamb-Dicke
regime). Actually --- both for `pedagogical reasons' and for a
wider comparison with the literature on this subject --- with
respect to sect.~{\ref{ion}}, we will consider a slightly more
general model. As in sect.~{\ref{ion}}, the Hilbert space of our
model is $\mathcal{H}_{\mathrm{F}}\otimes\,\mathbb{C}^2$ where
$\mathcal{H}_{\mathrm{F}}$ is the Fock space, namely a
infinite-dimensional Hilbert space endowed with an orthonormal
basis $\{\left|n\right\rangle: n=0,1,\ldots\}$, and with the
annihilation and creation operators $a,\,a^\dagger$ associated
with this basis: $a\left|0\right\rangle=0\,,\ a
\left|n\right\rangle=\sqrt{n}\,\left|n-1\right\rangle ,\
n=1,2,\ldots\ $. The Hamiltonian that we will consider in this
section is of the form
\begin{equation}
\hlt=H_0 + \hilt\,,
\end{equation}
where the unperturbed component $H_0$, as in sect.~{\ref{ion}}, is given
by\footnote{In this section, with respect to sect.~{\ref{ion}}, we will
adopt a somewhat more formal notation that highlights the tensor product structure
of our model.}
\begin{equation}
H_0=\nu \; \hat{n}\otimes\mathrm{Id}_{\mbox{\tiny $\mathbb{C}^2$}}
+\frac{1}{2}\,\epsilon\;\mathrm{Id}_{\mbox{\tiny
$\mathcal{H}_{\mathrm{F}}$}}\!\otimes\,\sigma_z\, ,\ \ \ \ \
\nu,\epsilon>0
\end{equation}
--- with $\hat{n}$ and $\sigma_z$ denoting respectively
the number operator $a^\dagger a$,
$\hat{n}\left|n\right\rangle=n\left|n\right\rangle$, and the
effective spin operator (associated with the internal degrees of
freedom of the two-level ion)
$\sigma_z\left|\pm\right\rangle=\pm\left|\pm\right\rangle,\
\left|+\right\rangle\equiv (1,0)\,,\;\left|-\right\rangle\equiv
(0,1)$ --- while the analytic perturbation $\hilt$ is now defined
by (compare with the interaction term $H_\sapprox(t)$ defined
by~{(\ref{appham})}):
\begin{equation}
\!\!\hilt := \lambda\nu\Big(\! e^{i\alpha t}\! \left(\gn + \eif(\af +
\afd)\right)\!\otimes\sigma_{-} + \, e^{-i\alpha t}\!\left(\gn +
\emif(\af + \afd)\right)\!\otimes\sigma _{+}\!\Big),
\end{equation}
with $\sigma_{\pm}=\left|\pm\right\rangle\left\langle\mp\right|\,$,\, $\sigma_+=\sigma_-^\dagger$
(as usual), with $\af$ denoting the `deformed oscillator operator' in
$\mathcal{H}_{\mathrm{F}}$ defined by
\begin{equation} \label{defdef}
\af:=a\,f(\hat{n})=f(\hat{n}+1)\,a\,,
\end{equation}
and with the functions
$g,f:\{0\}\cup\mathbb{N}\rightarrow\mathbb{R}$ (notice that, due
to~{(\ref{defdef})}, we can set, without loss of generality,
$f(0)\equiv 1$) and the phase factor $\eif$ characterizing the
model. Observe that, since the functions $g,f$ are assumed to be
$\mathbb{R}$-valued, we have:
\begin{equation}
\gn^\dagger=\gn\, ,\ \ \ f(\hat{n})^\dagger=f(\hat{n})\, ,\ \ \
\afd=f(\hat{n})\,a^\dagger=a^\dagger f(\hat{n}+1)\, .
\end{equation}
Obviously, setting
\[
g(n)=1\, ,\ \  f(n+1)=\eta\,, \ \ \forall
n\in\{0\}\cup\mathbb{N}\, ,\ \ \ \eif=-i\, ,
\]
we recover the model, with interaction term $H_\sapprox(t)$,
considered in sect.~{\ref{ion}}; while if we set instead
$\epsilon=\nu$, $\alpha=0$ and $g(n)=0$, $f(n+1)=\sqrt{n+1},\
\forall n\in\{0\}\cup\mathbb{N}$, $\,e^{i\phi}=1$, we obtain a
generalization with `counter-rotating terms' (i.e.
$\lambda\nu\,\af\otimes\sigma_-$ and
$\lambda\nu\,\afd\otimes\sigma_+$) of the `Jaynes-Cummings model
with intensity-dependent coupling' studied,
for instance, in ref.~\cite{Chaichian}.\\
Another natural choice of the functions $g,f$ is provided by a
more accurate approximation of the interaction term $H_\updown
(t)$ of the ion trap Hamiltonian, approximation which is adopted
by some authors (see, for instance,
refs.~{\cite{Blockey,Vogel,Vogel*}}). Indeed, it turns out that we
have:
\begin{eqnarray}
H_\updown(t) \spa &=& \spa \lambda\,\nu\left( e^{i\alpha t}\,
D(-i\eta)\,\sigma_- + e^{-i\alpha t}\, D(i\eta)\,\sigma_+\right)
\nonumber \\
& = & \spa \lambda\,\nu \bigg(
e^{i\alpha t}\, e^{-\eta^2/2} \Big(\fun_0(\eta;\hat{n})
\nonumber\\
& + & \spa \sum_{m=1}^\infty (-i \eta)^m \big(\fun_m(\eta;\hat{n})\,
a^m + a^{\dagger} {}^m \fun_m(\eta;\hat{n})\big) \Big)\otimes \sigma_- + \, h.c.
\bigg),
\end{eqnarray}
where the operator functions $\fun_0(\eta;\hat{n}),\,
\fun_m(\eta;\hat{n}),\,m=1,2,\ldots\;$, are defined by
\begin{eqnarray}
\fun_0(\eta;\hat{n}) \spa & := & \spa \sum_{l=0}^\infty
\frac{(i\,\eta)^{2l}}{(l!)^2}\, (a^\dagger)^l a^l =
\sum_{l=0}^\infty \frac{(i\,\eta)^{2l}}{(l!)^2}\, [\hat{n}]_l\, ,
\label{fiz}
\\
\fun_m(\eta;\hat{n}) \spa & := & \spa \sum_{l=0}^\infty
\frac{(i\,\eta)^{2l}}{l!\,(l+m)!} \;[\hat{n}]_l\, , \label{fim}
\end{eqnarray}
with:
\begin{eqnarray}
[n]_0\equiv 1\, ,\spa & & [n]_l\equiv n\,(n-1)_+\cdots\,(n-l+1)_+\, ,\ \
l\ge 1\, ,\nonumber\\
(n-m)_+\!\equiv 0 \ \ \ \mbox{for}\ n<m\, , \spa  & & (n-m)_+\!\equiv
n-m \ \ \ \mbox{for}\ n\ge m\, .
\end{eqnarray}
Notice that the functions $\{0\}\cup\mathbb{N}\ni
n\mapsto\fun_k(\eta;n)$, $\,k=0,1,\ldots\;$, are
$\mathbb{R}$-valued.\\ In order to derive this result, it is
sufficient to observe that
\begin{eqnarray}
\spa\spa D(\pm i\eta) \spa & := & \spa
\exp\!\left(\pm i\,\eta\, (a+a^\dagger)\right)  \nonumber \\
& = & \spa
e^{-\eta^2/2}\; e^{\pm i\,\eta\,a^\dagger}\, e^{\pm i\,\eta\,a}
\nonumber\\
& = & \spa e^{-\eta^2/2}\, \bigg(\,\sum_{l=0}^\infty
\frac{(i\,\eta)^{2l}}{(l!)^2}\ \overbrace{\underbrace{\,a^\dagger
\cdots\, a^\dagger\!}_l\ \underbrace{\,a\,\cdots\, a\,}_l}^{\
[\hat{n}]_l}
\nonumber\\
& \pm & \spa \! i\,\eta \sum_{l=0}^\infty
\frac{(i\,\eta)^{2l}}{l!(l+1)!}\,
\big(\underbrace{\,a^\dagger\cdots\,a^\dagger\!}_l\
\underbrace{\,a\,\cdots\,a\,}_l\; a
+a^\dagger\underbrace{\,a^\dagger\cdots\,a^\dagger\!}_l\
\underbrace{\,a\,\cdots\,a\,}_l\big)
+ \cdots \bigg),
\end{eqnarray}
from which formulae~{(\ref{fiz})} and~{(\ref{fim})} are readily obtained.
Notice also that, recalling the expression of the generalized
Laguerre polynomial
\begin{equation}
L_n^m(z)=\sum_{l=0}^n \binom{n+m}{n-l}\frac{(-z)^l}{l!}\,,
\end{equation}
we have:
\begin{equation}
\fun_m(\eta;n)=\frac{n!}{(m+n)!}\;L_{n}^m(\eta^2)\,,\
\ \ \ m,n=0,1,2,\ldots\ .
\end{equation}
Then, in the Lamb-Dicke regime ($\eta\ll1$), we can set (recall
that $f(0)\equiv 1$):
\[
g(n)=e^{-\eta^2/2}\,\fun_0(\eta;n)\, ,\ \  f(n+1)=\eta\;
e^{-\eta^2/2}\,\fun_1(\eta;n)\, ,\ \ \forall
n\in\{0\}\cup\mathbb{N}\, , \ \ \ \eif=-i\, .
\]

We are not interested here in performing a complete analysis of the dynamics
generated by the Hamiltonian $\hlt$, analysis which would involve a
careful study
of the different regimes associated with specific ranges of the parameters
$\nu,\epsilon$ and $\alpha$. What is of interest to us is to fix these parameters
in such a way that most of the relevant features of our perturbative expansion
can be illustrated. We will assume, then, that the
parameters $\nu,\alpha,\epsilon$ satisfy the resonance condition
\begin{equation} \label{resonance}
\delta\equiv\epsilon-\alpha=\nu \, ,
\end{equation}
which is also the most interesting regime from the point
of view of applications. With condition~{(\ref{resonance})},
the interaction picture Hamiltonian has the following expression:
\begin{eqnarray}
\rothamlt \spa & = & \spa \exp(i\ho t)\,\hilt\,\exp(-i\ho t)
\nonumber\\
& = & \spa \lambda\nu\Big(e^{i(\alpha -\epsilon)t}\, \gn\,
\otimes\,\sigma_{-} +\, \eif (e^{i(\alpha -\nu-\epsilon)t}\, \af
\otimes\,\sigma_{-} + \, e^{-i(\alpha-\nu+\epsilon)t}\,
\afd\otimes\,\sigma _{-})+\, h.c.\Big) \nonumber\\ & = & \spa
\lambda\nu\Big(e^{-i\nu t}\, \gn\, \otimes\,\sigma_{-} + \eif(
e^{-i2\nu t}\, \af\otimes \,\sigma_{-} + \; \afd\otimes\,\sigma
_{-}) +\, h.c.\Big)\,.
\end{eqnarray}
Notice that it depends periodically on time, with period $\period=2\pi/\nu$.\\
At this point, we recall that there is an almost ubiquitous
empirical rule in quantum optics,
the rotating wave approximation --- aimed to drastically simplify the
determination of the evolution operator
--- that says:
\begin{quote}
{\em ``in order to compute the evolution operator, skip the rapidly oscillating terms
in the interaction picture Hamiltonian''}.
\end{quote}
This recipe, applied to our case, would lead us to consider for
the interaction picture Hamiltonian the following approximation:
\begin{equation} \label{rwapp}
\rotham\lt\approx \lambda\,\nu\,(\eif\,
\afd\otimes\,\sigma_-+\,\emif\,\af\otimes\,\sigma_+) =: \eff(\lambda)\, .
\end{equation}
Clearly, the effective Hamiltonian $\eff(\lambda)$ --- with
$f(n)=\eta\, ,\ n=0,1,\ldots\;$, and $\eif=-i$ --- coincides with the
effective Hamiltonian $\heffp$ found in sect.~{\ref{ion}}.
We are going to show that the `approximation'~{(\ref{rwapp})} does not even produce
the correct first order expression of the evolution operator.

To obtain the first order perturbative expression of the evolution
operator associated with the Hamiltonian $\hlt$, we will first
exploit the method outlined in sect.~{\ref{generalcase}}.  To this aim,
it will be convenient to introduce
the analytic function $\avxp:\mathbb{C}\rightarrow\mathbb{C}$, with
\begin{equation}
\avxp(z)=\frac{e^z-1}{z} \ \ \ \mbox{for}\ z\neq 0\,,\ \ \
\avxp(0)=1\, ,
\end{equation}
and to define, for $\tau > 0$, the operator
\begin{eqnarray}
\tauc_1 \spa & := & \spa \frac{1}{\lambda\tau}\int_0^\tau
\rothamlt\ \de t
\nonumber\\
& = & \spa \nu \Big(\avxp(-i\nu\tau)\;\gn\,\otimes\,\sigma_-
\nonumber \\
& + & \spa \eif\big(\avxp(-i2\nu\tau)\;\af\otimes\,\sigma_- +
\,\afd\otimes\,\sigma_-\big) +\, h.c. \Big)\,.
\end{eqnarray}
From this expression, we have immediately (recall
formulae~{(\ref{limits})}, and notice that, in this case,
$\rothamlt=\lambda\,\rotham_1(t)$) that the operator $\infc_1$ is
given by
\begin{equation} \label{op1}
\infc_1 = \lim_{\tau\rightarrow\infty} \tauc_1 = \nu \left(\eif\,
\afd\otimes\,\sigma_- + \,\emif\,\af\otimes\,\sigma_+ \right)\,.
\end{equation}
Then, we can easily obtain the operator $\infz_1$:
\begin{eqnarray}
\infz_1 \spa & = & \spa \lim_{\tau\rightarrow\infty}
\left(\frac{1}{2}\tau\,\infc_1
-\frac{1}{\tau}\int_0^\tau t\ \tc_1\,\de t\right)\nonumber \\
& = & \spa i\,\gn\otimes(\sigma_- - \, \sigma_+
)+\frac{i}{2}\left(\eif\,\af\otimes\,\sigma_-
-\,\emif\,\afd\otimes\,\sigma_+\right). \label{op2}
\end{eqnarray}
Next, the expression of the operator-valued function $t\mapsto
Z_1(\infc_1,\infz_1;t)$ is found to be
\begin{eqnarray}
Z_1(\infc_1,\infz_1;t) \spa & = & \spa \infz_1 +t\left(\tc_1-
\infc_1 \right)\nonumber \\
& = & \spa i\,\gn\otimes\left(e^{-i\nu t}\,\sigma_- - \,e^{i\nu
t}\, \sigma_+\right) \nonumber
\\
& + & \spa \frac{i}{2} \left( e^{-i(2\nu
t-\phi)}\,\af\otimes\,\sigma_- -\, e^{i(2\nu
t-\phi)}\,\afd\otimes\,\sigma_+\right); \label{op3}
\end{eqnarray}
hence, it is periodic on time, with the same period
$\period=2\pi/\nu$ of the operator-valued function
$t\mapsto\rotham\lt$. The reader may verify that the second order
contributions can also be calculated with a rather modest
effort.\\
We notice explicitly that formula~{(\ref{op1})} shows that the
operator $\lambda\,\infc_1$ coincides with the expression of the
effective interaction picture Hamiltonian prescribed by the rotating wave
approximation, but, due to formulae~{(\ref{op2})}
and~{(\ref{op3})}, this prescription does not provide a correct
first order approximate expression of the evolution operator (we will
return to this point at the end of the section).
Besides, observe that, since $\infz_1\neq
0$, we are {\it not} recovering the Floquet-Magnus expansion.\\
We stress that we could have obtained this result also by looking
for the first order correction to the evolution operator
associated with the effective Hamiltonian $\eff(\lambda)$ --- as
explained in sect.~{\ref{generalcase}} --- with the only
additional requirement that the mean value of the function
$t\mapsto Z_1(\infc_1,\infz_1;t)$ be zero (condition that fixes
the arbitrary constant $\infz_1$).

Eventually, the unitary operators generated by the selfadjoint
operators~{(\ref{op1})}, (\ref{op2}) and~{(\ref{op3})} can be
explicitly computed. Indeed, using the fact that
\begin{eqnarray}
\spa\spa\spa\left(e^{i\theta}\,\af\otimes\,\sigma_\pm \pm
\,e^{-i\theta}\,\afd\otimes\,\sigma_\mp\right)^{2m} \!\!\! \spa &
= & \spa \left(\pm\,(\afd\, \af)\otimes (\sigma_\mp\,
\sigma_\pm)\pm  (\af \,\afd)\otimes (\sigma_\pm\,\sigma_\mp)
\right)^m
\nonumber \\
& = & \spa (\pm 1)^m \left(\effe(\hat{n})^{2m}
\left|\mp\right\rangle\left\langle \mp\right| +
\effe(\hat{n}+1)^{2m}\left|\pm\right\rangle\left\langle
\pm\right|\right)  \label{tool}
\end{eqnarray}
--- where we have set, for the sake of notational conciseness,
\begin{equation}
\effe(n):=f(n)\,\sqrt{n} \ \ \ \mbox{for}\ n\in\mathbb{N}\,,\ \ \
\effe(0)\equiv 1
\end{equation}
--- we find easily:\footnote{In the following we will set, by
convention, $\sin(0)/0\equiv 1$.}
\begin{eqnarray}
\spa\spa\spa\spa\!\exp(-i\lambda\,\infc_1\, t) \spa & = & \spa
\exp\!\left(-i\lambda\nu\big(
\afd\otimes\,\sigma_- + \,\af\otimes\,\sigma_+ \big)\, t\right) \nonumber\\
& = & \spa \cos\left(\lambda\nu\,\effe(\hat{n})\;
t\right)\,\otimes\, \left|-\right\rangle\left\langle -\right| +\,
\cos\left(\lambda\nu\,\effe(\hat{n}+1)\;
t\right)\,\otimes\, \left|+\right\rangle\left\langle +\right|
\nonumber \\
& - & \spa i\!\left(\eif\,\mbox{\small
$\dfrac{\sin\left(\lambda\nu\,\effe(\hat{n})\;
t\right)}{\effe(\hat{n})}$}\, \afd\otimes\,\sigma_-
+\,\emif\,\mbox{\small
$\dfrac{\sin\left(\lambda\nu\,\effe(\hat{n}+1)\; t\right)
}{\effe(\hat{n}+1)}$}\, \af\otimes\, \sigma_+\right) .
\end{eqnarray}
Next, observe that
\begin{equation} \label{formprod}
e^{-i\,\lambda\, Z_1(\infc_1,\infz_1;t)}\,\appr\,
e^{\lambda\,\gn\otimes(e^{-i\nu t}\,\sigma_- - \;e^{i\nu t}\,
\sigma_+ )}\; e^{\frac{1}{2}\,\lambda \left( e^{-i2\nu
t}\,\af\otimes\,\sigma_- -\, e^{i2\nu
t}\,\afd\otimes\,\sigma_+\right)},
\end{equation}
hence, it will be sufficient to compute the separate exponentials that
appear on the r.h.s.\ of eq.~{(\ref{formprod})}. The computation of the
first exponential is straightforward:
\begin{eqnarray}
e^{\lambda\,\gn\otimes(e^{-i\nu t}\,\sigma_- - \; e^{i\nu
t}\,\sigma_+ )} \spa & = & \spa
\cos(\lambda\,\gn)\otimes\mathrm{Id}_{\mbox{\tiny $\mathbb{C}^2$}}
+\,\sin(\lambda\,\gn)\otimes(e^{-i\nu t}\,\sigma_- -\;e^{i\nu t}\,\sigma_+)
\nonumber \\
& =: & \spa \wuuno\lt\,. \label{vuuno}
\end{eqnarray}
Then, using again relation~{(\ref{tool})}, we find:
\begin{eqnarray}
e^{\frac{1}{2}\,\lambda \left( e^{-i2\nu t}\,\af\otimes\,\sigma_-
-\, e^{i2\nu t}\,\afd\otimes\,\sigma_+\right)} \spa & = & \spa
\cos\left(\mbox{$\frac{1}{2}$}\lambda\,\effe(\hat{n}+1)\right)\otimes\,
\left|-\right\rangle\left\langle -\right|\, +\;
\cos\left(\mbox{$\frac{1}{2}$}\lambda\,\effe(\hat{n})\right)\otimes\,
\left|+\right\rangle\left\langle +\right|
\nonumber \\
& + & \spa \bigg( e^{-i(2\nu t-\phi)}\,\mbox{\small
$\dfrac{\sin\!\left(\frac{1}{2}\lambda\,\effe(\hat{n}+1)
\right)\!}{\effe(\hat{n}+1)}$}\, \af\otimes\,\sigma_-\!
\nonumber\\
& - & \spa e^{i(2\nu t-\phi)}\,\mbox{\small
$\dfrac{\sin\!\left(\frac{1}{2}\lambda\,\effe(\hat{n})
\right)\!}{\effe(\hat{n})}$}\, \afd\otimes\,
\sigma_+\bigg)=:\wudue\lt\, . \label{vudue}
\end{eqnarray}
From formulae~{(\ref{vuuno})} and~{(\ref{vudue})}, we can obtain immediately
the expression of $\exp(i\lambda\,\infz_1)$; indeed:
\begin{eqnarray}
\exp(i\lambda\,\infz_1)\spa & = & \spa \exp\!\left(i\lambda\,
Z_1(\infc_1,\infz_1;0)\right)
\nonumber \\
& \appr & \spa e^{-\frac{1}{2}\,\lambda \,( \af\otimes\,\sigma_-
-\; \afd\otimes\,\sigma_+)}\  e^{-\lambda\,\gn\otimes(\sigma_- -
\; \sigma_+ )}
\nonumber \\
& = & \spa \wudue(\lambda;0)^\dagger\, \wuuno (\lambda;0)^\dagger
= \,  \wudue(-\lambda;0)\; \wuuno (-\lambda;0)\, .
\end{eqnarray}
The exponentials above provide a simple explicit form of the first order
approximation of the evolution operator:
\begin{eqnarray}
\rotev (\lambda ;t) \spa & \appr & \spa
\exp\!\left(-i\lambda\,Z_1(\infc_1,\infz_1;t)\right)\,\exp(-i\lambda\,\infc_1\,
t) \,\exp\! \left(i\lambda\,\infz_1\right) \nonumber\\
& \appr & \spa \wuuno\lt\;\wudue\lt\; \exp(-i\lambda\,\infc_1\, t)
\ \wudue(-\lambda;0)\; \wuuno (-\lambda;0)\,.
\end{eqnarray}
It is worth stressing the non-trivial fact that
the action of this operator on the standard basis
$\{\left|n\right\rangle\otimes\left|\pm\right\rangle :
n=0,1,\ldots\}$ can be easily computed.

As already observed in sect.~{\ref{ion}}, a remarkable feature of
the time-dependent Hamiltonian $\hlt$ is that the associated
dynamics can be transformed into the dynamics generated by a
time-independent Hamiltonian by switching to a suitable
interaction picture, i.e.\ the interaction picture determined by
the reference Hamiltonian $\frac{1}{2}\,\alpha
\;\idf\!\otimes\,\sigma_{z}$. Indeed, setting
\begin{equation}
R_t^{} := \exp\!\left(-\frac{i}{2}\,\alpha\;\idf\!\otimes
\,\sigma_z\; t \right),
\end{equation}
one can define the time-independent `rotating frame Hamiltonian'
\begin{equation}
\mathfrak{H}(\lambda) := R_{t}^{\dagger}\left(\hlt-\frac{1}{2}\,
\alpha\;\idf\!\otimes\,\sigma_{z}\right) R_{t}^{} = \mathfrak{H}_0
+ \mathfrak{H}_{\diamond}(\lambda)\, ,
\end{equation}
where (recall relation (\ref{resonance})):
\begin{eqnarray}
\mathfrak{H}_0 \spa & :=  & \spa \nu\,\hat{n}\otimes\,\idc
+\frac{1}{2}\,(\epsilon-\alpha)\;\idf\!\otimes\,\sigma_{z}
=\nu\left(\hat{n}\otimes\,\idc+\frac{1}{2}\;\idf\!\otimes\,\sigma_z\right), \\
\spa\spa \mathfrak{H}_{\diamond}(\lambda) \spa & := & \spa
\lambda\,\nu \Big( \left(\gn + \eif(\af +\,
\afd)\right)\otimes\,\sigma_- +\,\left(\gn +
\emif(\af+\,\afd)\right)\otimes\, \sigma_+ \Big)\, .
\end{eqnarray}
We notice explicitly that the further interaction picture
Hamiltonian $\tilde{\mathfrak{H}}\lt$, obtained from the
time-independent Hamiltonian $\mathfrak{H}(\lambda)$ taking as
reference Hamiltonian the unperturbed component $\mathfrak{H}_0$,
coincides with the `old' interaction picture Hamiltonian
$\rotham\lt$; indeed:
\begin{eqnarray}
\tilde{\mathfrak{H}}\lt \spa & := & \spa e^{i\mathfrak{H}_0 t}\
\mathfrak{H}_{\diamond}(\lambda) \; e^{-i\mathfrak{H}_0 t}
\nonumber \\
& = & \spa \lambda\nu\Big(e^{-i\nu t}\, \gn \otimes \,\sigma_{-} +
\, \eif(e^{-i2\nu t}\, \af\otimes \,\sigma_{-} + \;
\afd\otimes\,\sigma _{-})+\, h.c.\Big) \nonumber \\
& = & \spa \rotham\lt\,.
\end{eqnarray}
Hence, we can refer, without ambiguity, to the operators
$\infc_1$, $\infz_1$, and $Z_1(\infc_1,\infz_1;t)$. At this point,
the reader may easily verify the following facts:
\begin{itemize}
\item using the results of sect.~{\ref{independent}}, one finds
that the operators $\minc_1$ and $\minz_1$ coincide respectively
with the operators $\infc_1$ and $\infz_1$ as given by
formulae~{(\ref{op1})} and~{(\ref{op2})};
\item the operator-valued function $t\mapsto
Z_1(\infc_1,\infz_1;t)$ satisfies the relation
\begin{equation}
Z_1(\infc_1,\infz_1;t)= e^{i\mathfrak{H}_0 t}\ \infz_1\;
e^{-i\mathfrak{H}_0 t}\,,
\end{equation}
in agreement with what has been stated in
sect.~{\ref{generalcase}}.
\end{itemize}
The link between the two descriptions --- the one associated with
the time-dependent Hamiltonian $H\lt$ and the one associated with
the time-independent Hamiltonian $\mathfrak{H}(\lambda)$ --- is
given by:
\begin{eqnarray*}
U\lt = R_{t}^{}\; e^{-i\mathfrak{H}(\lambda)t} \spa & \appr & \spa
R_{t}^{}\;e^{-i\lambda\minz_1}\,e^{-i(\mathfrak{H}_0+\lambda\minc_1)
t}\, e^{i\lambda\minz_1}
\\
\spa & = & \spa
R_{t}^{}\;e^{-i\lambda\infz_1}\,e^{-i(\mathfrak{H}_0+\lambda\infc_1)
t}\, e^{i\lambda\infz_1} \\
& = & \spa R_{t}^{}\; e^{-i\mathfrak{H}_0
t}\left(e^{i\mathfrak{H}_0 t}\,
e^{-i\lambda\infz_1}\,e^{-i\mathfrak{H}_0 t} \right)
e^{-i\lambda\infc_1 t}\, e^{i\lambda\infz_1}
\\
& = & \spa e^{-i\ho t}\,e^{-i\lambda \,
Z_1(\infc_1,\infz_1;t)}\,e^{-i\lambda\infc_1 t}\,
e^{i\lambda\infz_1} \\
& \appr & \spa e^{-i\ho t}\, \rotev\lt = U\lt\, .
\end{eqnarray*}

We conclude this section with a final comment. Consider the
structure of the first order expression of the evolution operator
generated by $\mathfrak{H}(\lambda)$; namely:
\[
\exp(-i\mathfrak{H}(\lambda)\,t)  \appr\,
\underbrace{\,\exp(-i\lambda\infz_1)}_{W(\lambda)}\
\underbrace{\,\exp(-i(\mathfrak{H}_0+\lambda\infc_1)\,
t)}_{\mathfrak{R}\lt}\
\underbrace{\,\exp(i\lambda\infz_1)}_{W(\lambda)^\dagger}.
\]
It is then evident that --- since $\mathfrak{R}\lt$ coincides with
the evolution obtained by the RWA --- the {\it true} first order
evolution operator can be expressed as the result of the
application of a time-independent unitary transformation (associated
with the counter-rotating terms) to the
RWA evolution. Hence, we can say that the RWA provides at least
the first order {\it qualitative behavior} of the evolution
operator associated with $\mathfrak{H}(\lambda)$. On the other
hand, observe that
\begin{eqnarray*}
\exp(-i\mathfrak{H}(\lambda)\,t) \spa & \appr& \spa
W(\lambda)\,\exp(-i\mathfrak{H}_0\,
t)\,W(\lambda)^\dagger\,\exp(-i \lambda\infc_1\, t)\\
& = & \spa \underbrace{(W(\lambda)\,\exp(-i\mathfrak{H}_0\,
t)\,W(\lambda)^\dagger \exp(i\mathfrak{H}_0\, t
))}_{\mathfrak{D}\lt} \;\mathfrak{R}\lt\, ,
\end{eqnarray*}
where, as it can be easily checked, the operator $\mathfrak{D}\lt$
gives a non-trivial first order contribution. Hence, the RWA does
not provide, already at the first perturbative
order, a correct approximation of the evolution operator associated with $\mathfrak{H}(\lambda)$.\\
It is also worth observing that the RWA and counter-rotating terms
play different roles --- but with the same `dignity' --- in the first order
approximation provided by our perturbative expansion of the evolution operator.


\section{Conclusions and just a glance to further applications}
\label{conclusions}

In the present paper, we have introduced a perturbative expansion
of the evolution operator associated with a, in general time-dependent,
quantum Hamiltonian that exploits the power of the perturbative approach
for a twofold purpose:
\begin{itemize}

\item to obtain {\it unitary} approximate expressions of the evolution operator,
as in the Magnus expansion, of which it is a generalization;

\item simultaneously, to achieve computational advantages, in the same
spirit of standard perturbation theory for linear operators.

\end{itemize}
On our opinion, it is a remarkable fact that the time-independent
perturbative approach (essentially, Rayleigh-Schr\"odinger-Kato
perturbation theory) and the time-dependent perturbative approach
(Dyson and Magnus expansions) --- that are traditionally regarded as
completely distinct subjects --- can be combined to form such a
non-trivial blend. This feature of our perturbative expansion
makes it possible, for instance, to treat closely related models ---
that may be described, depending on the particular case,
by a time-independent or by a time-dependent Hamiltonian --- using
the same general method, so that a direct comparison of the results
obtained is achievable.

Remarkable examples of applications of the methods outlined in the
present paper are the applications to quantum optical systems. In
particular, applications to the study of laser-driven trapped ions
(or ions in optical cavities) with various coupling schemes are
likely to be very fruitful~\cite{Aniello1}~\cite{Militello}, also
in view of the possible implementation, by means of such devices,
of quantum computers~\cite{Nielsen}. For some of these systems,
one can switch to a suitable interaction picture to obtain a
time-independent effective Hamiltonian, as we have seen in the
example studied in sects.~{\ref{ion}} and \ref{applications}, and
apply the results of sect.~\ref{independent}. In some other cases,
this trick does not work and one can apply the more general
results of sect.~\ref{generalcase}; a simple example is provided
by the Hamiltonian describing a trapped ion in interaction with a
bichromatic laser field. In all the mentioned cases, the results
presented here allow to overcome the severe limitations imposed by
the {\it rotating wave approximation} usually adopted in the
literature.

Although a careful criticism of the RWA is not the subject of this
paper, it is worth mentioning briefly a few facts. As already
observed long time ago by Agarwal~{\cite{Agarwal1, Agarwal2}}, who
studied spontaneous emission effects in two-level ions, the RWA
{\it should not} be applied directly to the Hamiltonian (hence,
for obtaining approximate expressions of the evolution operator),
but only --- possibly and with due caution ---
in the calculation of some experimentally
observable quantities; if one insists in applying the RWA to the
Hamiltonian, the agreement with the `true' behavior of the experimentally
observable quantities can strongly depend both on the choice of the
observable itself and on the initial condition of the system. The main
point to understand is that the behavior of a single observable
quantity does not contain, in general, all the `information' encoded in the
evolution operator.\\
Several authors have investigated the contribution of the
`counter-rotating terms' in various quantum optical models,
contribution which is completely ignored by the RWA; for instance,
perturbative corrections to the energy spectrum~\cite{Cohen}, and
corrections to the time evolution by means of path
integral~\cite{Zaheer}, perturbative~\cite{Vyas, Phoenix, Fang} and
numerical~\cite{Seke, Berlin}
techniques have been studied. However, as far as we know, our
approach is the first systematic attempt at taking into account
the corrections to the RWA evolution operator from a completely general
point of view, which allows also to gain a deeper insight in the
different roles played by the RWA and counter-rotating terms.

We conclude this section with a final comment. The method
introduced in the present paper can be extended in a
straightforward way to obtain a perturbative expansion of the {\it
propagator} (or `evolution family', i.e.\ the generalization of
the unitary evolution operator) associated with a perturbed {\it
evolution equation} in a Banach space (a classical reference on,
in general {\it temporally inhomogeneous} or `non-autonomous',
evolution equations is~\cite{Yosida}; see also ref.~\cite{Engel}).
Indeed, our method relies on a pure operator approach. This fact
implies that it could be applied, in particular, to the study of
the dynamics of {\it open quantum
systems}~\cite{Gorini,Lindblad,Holevo,Petruccione}, namely, of
quantum systems subject to a coupling with an uncontrollable
environment or `bath', coupling that is responsible for the
phenomenon known as {\it quantum decoherence}. Under certain
physical conditions, the dynamics of such a system can be
described by a propagator whose infinitesimal generator is a
`superoperator' acting on the density operators, i.e.\ on the
positive unit trace operators in the Hilbert space of the quantum
system (that form a convex subset of the Banach space of trace
class operators). In many important cases (see
ref.~\cite{Petruccione} and the rich bibliography therein), the
`dissipative component' of the superoperator can be considered as
(part of) a perturbation, hence, the associated propagator may be
expanded using our method.

\section*{Acknowledgements}

\noindent The main results of the paper were presented by the
author at the {\it 9-th International Conference on Squeezed
States and Uncertainty Relations}, held in Besan\c{c}on (France),
May 2--6, 2005. The author wishes to thank the organizers for the
very kind hospitality.

The author wishes to dedicate the present paper, with deep admiration
and gratitude,
to Prof.\ F. Zaccaria, a bright example of a scientist and a wonderful person.


\end{document}